\documentclass[12pt]{article}
\usepackage{amsmath}
\usepackage{graphicx}
\usepackage{natbib}
\usepackage{url} 

\newcommand{\blind}{0}

\usepackage[T1]{fontenc}
\usepackage[utf8]{inputenc}
\usepackage{lmodern}

\usepackage{geometry}
\usepackage{amsmath}
\usepackage{amssymb}
\usepackage{array}
\usepackage{mathtools}

\usepackage{multirow}

\usepackage{algorithm}

\newtheorem{assumption}{Assumption}

\usepackage{graphicx}

\usepackage{tikz}
\usetikzlibrary{fit}
\tikzset{
  randomnode/.style={draw, minimum width=2em, shape=circle, inner sep=0em},
  fixednode/.style={draw, minimum width=2em, minimum height=2em, shape=rectangle, inner sep=0em},
  dagconn/.style={arrows=->, >=stealth},
  plate/.style={draw, shape=rectangle, rounded corners=0.5em, minimum width=7em, minimum height=9em, anchor=center}
}

\usepackage{caption}
\usepackage{subcaption}

\newcommand{\reals}{\mathbb{R}}
\newcommand{\posreals}{\reals_{+}}
\newcommand{\sfE}{\mathsf{E}}

\newcommand{\E}{\operatorname{\mathbb{E}}}
\newcommand{\var}{\operatorname{var}}

\newcommand{\diff}{\mathop{}\!\text{d}}

\newcommand{\Id}{\operatorname{Id}}

\newcommand{\dnorm}{\mathcal{N}}
\newcommand{\dlnorm}{\mathcal{LN}}

\begin{document}

\def\spacingset#1{\renewcommand{\baselinestretch}%
{#1}\small\normalsize} \spacingset{1}


\if0\blind
{
  \title{\bf Global Consensus Monte Carlo}
  \author{Lewis J.\ Rendell\thanks{
    The authors gratefully acknowledge the support of The Alan Turing Institute under the EPSRC grant EP/N510129/1, the Lloyd's Register Foundation--Alan Turing Institute Programme on Data-Centric Engineering, and the EPSRC under grants EP/M508184/1, EP/R034710/1 and EP/T004134/1.}\hspace{.2cm}\\
    Department of Statistics, University of Warwick\\
    and \\
    Adam M.\ Johansen \\
    Department of Statistics, University of Warwick \\
	and \\
	Anthony Lee \\
	School of Mathematics, University of Bristol \\
	and \\
	Nick Whiteley \\
	School of Mathematics, University of Bristol \\}
  \maketitle
} \fi

\if1\blind
{
  \bigskip
  \bigskip
  \bigskip
  \begin{center}
    {\LARGE\bf Global Consensus Monte Carlo}
\end{center}
  \medskip
} \fi

\bigskip
\begin{abstract}
To conduct Bayesian inference with large data sets, it is often convenient or necessary to distribute the data across multiple machines.
We consider a likelihood function expressed as a product of terms, each associated with a subset of the data.
Inspired by global variable consensus optimisation, we introduce an instrumental hierarchical model associating auxiliary statistical parameters with each term, which are conditionally independent given the top-level parameters.
One of these top-level parameters controls the unconditional strength of association between the auxiliary parameters.
This model leads to a distributed MCMC algorithm on an extended state space yielding approximations of posterior expectations. A trade-off between computational tractability and fidelity to the original model can be controlled by changing the association strength in the instrumental model.
We further propose the use of a SMC sampler with a sequence of association strengths, allowing both the automatic determination of appropriate strengths and for a bias correction technique to be applied.
In contrast to similar distributed Monte Carlo algorithms, this approach requires few distributional assumptions.
The performance of the algorithms is illustrated with a number of simulated examples.
\end{abstract}

\noindent%
\vfill

\newpage
\spacingset{1.5} 

\section{Introduction}

Large data sets arising in modern statistical applications present serious challenges for standard computational techniques for Bayesian inference, such as Markov chain Monte Carlo (MCMC) and other approaches requiring repeated evaluations of the likelihood function.
We consider here the situation where the data are distributed across multiple computing nodes.
This could be because the likelihood function cannot be computed on a single computing node in a reasonable amount of time, e.g.\ the data might not fit into main memory.

We assume that the likelihood function can be expressed as a product of terms so that the posterior density for the statistical parameter $Z$ satisfies
\begin{equation}
\label{eq:piOriginal}
\pi(z) \propto \mu(z) \prod_{j=1}^{b} f_{j}(z),
\end{equation}
where $Z$ takes values $z \in \sfE \subseteq \reals^{d}$, and $\mu$ is a prior density.
We assume that $f_{j}$ is computable on computing node $j$ and involves consideration of $\mathbf{y}_{j}$, the $j$th subset or `block' of the full data set, which comprises $b$ such blocks.

Many authors have considered `embarrassingly parallel' MCMC algorithms approximating expectations with respect to \eqref{eq:piOriginal}, following the Consensus Monte Carlo approach of \citet{ScottEtAl2016}.
Such procedures require separate MCMC chains to be run on each computing node, each targeting a distribution dependent only on the associated likelihood contribution $f_{j}$.
The samples from each of these chains are then combined in a final post-processing step to generate an approximation of the true posterior $\pi$.
Such algorithms require communication between the nodes only at the very beginning and end of the procedure, falling into the MapReduce framework \citep{DeanGhemawat2008}; their use is therefore more advantageous when inter-node communication is costly, for example due to high latency.
However, the effectiveness of such approaches at approximating the true posterior $\pi$ depends heavily on the final combination step.
Many proposed approaches are based on assumptions on the likelihood contributions $f_{j}$, or employ techniques that may be infeasible in high-dimensional settings.
We later review some of these approaches, and some issues surrounding their use, in Section~\ref{subsec:relatedApproaches}.

Instead of aiming to minimise entirely communication between nodes, we propose an approach that avoids employing a final aggregation step, thereby avoiding distributional assumptions on $\pi$.
This approach is motivated by global variable consensus optimisation \cite[see, e.g.,][Section~7; we give a summary in Section~\ref{subsec:instrumentalModel}]{BoydEtAl2011}.
Specifically we introduce an instrumental hierarchical model, associating an auxiliary parameter with each likelihood contribution (and therefore each computing node), which are conditionally independent given $Z$.
An additional top-level parameter controlling their unconditional strength of association is also introduced.
This allows the construction of an MCMC algorithm on an extended state space, yielding estimates of expectations with respect to $\pi$.
By tuning the association strength through the top-level parameter, a trade-off between computational tractability and fidelity to the original model can be controlled.

As well as avoiding issues associated with embarrassingly parallel algorithms, our framework presents benefits compared to the simple approach of constructing an MCMC sampler to directly target \eqref{eq:piOriginal}.
In settings where communication latency is non-negligible but the practitioner's time budget is limited, our approach allows a greater proportion of this time to be spent on computation rather than communication, allowing for faster exploration of the state space.

Our approach was initially presented in \citet{RendellEtAl2018}.
A proposal to use essentially the same framework in a serial context has been independently and contemporaneously published by \citet{VonoEtAl2019}, who construct a Gibbs sampler via a `variable splitting' approach.
Rather than distributing the computation, the authors focus on the setting where $b=1$ in order to obtain a relaxation of the original simulation problem.
An implementation of this approach for problems in binary logistic regression has been proposed in \citet{VonoEtAl2018}, with a number of non-asymptotic and convergence results presented more recently in \citet{VonoEtAl2019a}.
Our work focuses on distributed settings, providing a sequential Monte Carlo implementation of the framework that may be used to generate bias-corrected estimates.

We introduce the proposed framework and the resulting algorithmic structure in Section~\ref{sec:instrumentalModelMCMC}, including some discussion of issues in its implementation, and comparisons with related approaches in the literature.
We then introduce a sequential Monte Carlo implementation of the framework in Section~\ref{sec:SMCimplementation}.
Various simulation examples are presented in Section~\ref{sec:examples}, before conclusions are provided in Section~\ref{sec:conclusion}.

\section{The instrumental model and MCMC}
\label{sec:instrumentalModelMCMC}

For simplicity, we shall occasionally abuse notation by using the same symbol for a probability measure on $\sfE$, and for its density with respect to some dominating measure.
For the numerical examples presented herein, $\sfE \subseteq \reals^d$ and all densities are defined with respect to a suitable version of the Lebesgue measure.
We use the notation $x_{m:n} \coloneqq (x_m, \ldots, x_n)$ for arbitrary $x_m, \ldots, x_n$.
For a probability density function $\nu$ and function $\varphi$ we denote by $\nu(\varphi)$ the expectation of $\varphi(Z)$ when $Z \sim \nu$, i.e.\
\[
\nu(\varphi) \coloneqq \int \varphi(z) \nu(z) \diff z.
\] 

\subsection{The instrumental model}
\label{subsec:instrumentalModel}

The goal of the present paper is to approximate \eqref{eq:piOriginal}.
We take an approach that has also been developed in contemporaneous work by \citet{VonoEtAl2019}, although their objectives were somewhat different.
Alongside the variable of interest $Z$, we introduce a collection of $b$ instrumental variables each also defined on $\sfE$, denoted by $X_{1:b}$.
On the extended state space $\sfE \times \sfE^{b}$, we define the probability density function $\tilde{\pi}_{\lambda}$ by
\begin{equation}
\tilde{\pi}_{\lambda}(z,x_{1:b})
\propto
\mu(z) \prod_{j=1}^{b} K_{j}^{(\lambda)}(z,x_{j}) f_{j}(x_{j}),
\label{eq:jointZXdensity}
\end{equation}
where for each $j \in \{1, \dots, b\}$, $\{K_{j}^{(\lambda)} : \lambda \in \posreals\}$ is a family of Markov transition densities on $\sfE$.
Defining
\begin{equation}
f_{j}^{(\lambda)}(z) \coloneqq
\int_{\sfE} K_{j}^{(\lambda)}(z,x) f_{j}(x) \diff x,
\label{eq:smoothedLikelihood}
\end{equation}
the density of the $Z$-marginal of $\tilde{\pi}_{\lambda}$ may be written as
\begin{equation}
\pi_{\lambda}(z) \coloneqq
\int_{\sfE^b} \tilde{\pi}_\lambda(z,x_{1:b}) \diff x_{1:b}
\propto \mu(z) \prod_{j=1}^{b} f_{j}^{(\lambda)}(z).
\label{eq:smoothedPosterior}
\end{equation}
Here, we may view each $f_{j}^{(\lambda)}$ as a smoothed form of $f_{j}$, with $\pi_{\lambda}$ being the corresponding smoothed form of the target density \eqref{eq:piOriginal}.

The role of $\lambda$ is to control the fidelity of $f_{j}^{(\lambda)}$ to $f_{j}$, and so we assume the following in the sequel.
\begin{assumption}
\label{ass:lambda}
For all $\lambda > 0$, $f_{j}^{(\lambda)}$ is bounded for each $j \in \{1, \ldots, b\}$; and $f_{j}^{(\lambda)} \to f_{j}$ pointwise as $\lambda \to 0$  for each $j \in \{1, \ldots, b\}$.
\end{assumption}
For example, Assumption~\ref{ass:lambda} implies that $\pi_\lambda$ converges in total variation to $\pi$ by Scheff\'{e}'s lemma \citep{Scheffe1947}, and therefore $\pi_\lambda(\varphi) \to \pi(\varphi)$ for bounded $\varphi : \sfE \to \reals$.
Assumption~\ref{ass:lambda} is satisfied for essentially any kernel that may be used for kernel density estimation; here, $\lambda$ takes a similar role to that of the smoothing bandwidth.

On a first reading one may assume that the $K_{j}^{(\lambda)}$ are chosen to be independent of $j$; for example, with $\sfE = \reals$ one could take $K_{j}^{(\lambda)}(z,x) = \dnorm(x;z,\lambda)$.
We describe some considerations in choosing these transition kernels in Section~\ref{subsec:choosingTransitionDensities} of the supplement, and describe settings in which choosing these to differ with $j$ may be beneficial.

\begin{figure}
\begin{centering}
 \begin{tikzpicture}[x=1em,y=1em]
   \node[randomnode] (Zorig) at (0, 12.5) {$Z$};
   \node[fixednode] (Yorig) at (0, 4.8) {$\mathbf{y}_j$};
   \node[plate] at (0, 6) [label={[xshift=-6em,yshift=1.8em]south east:$j = 1, \dots, b$}] (plate1) {};
   \draw[dagconn] (Zorig) to (Yorig);

   \node[randomnode] (Zinst) at (10, 12.5) {$Z$};
   \node[fixednode] (lambda) at (14, 12.5) {$\lambda$};
   \node[randomnode] (X) at (12, 8.1) {$X_j$};
   \node[fixednode] (Yinst) at (12, 4.8) {$\mathbf{y}_j$};
   \node[plate] at (12, 6) [label={[xshift=-6em,yshift=1.8em]south east:$j = 1, \dots, b$}] (plate1) {};
   \draw[dagconn] (Zinst) to (X);
   \draw[dagconn] (lambda) to (X);
   \draw[dagconn] (X) to (Yinst);
 \end{tikzpicture}
\par\end{centering}
\caption{Directed acyclic graphs, representing the original statistical model (left) and the instrumental model we construct (right).}
\label{fig:DAGs}
\end{figure}
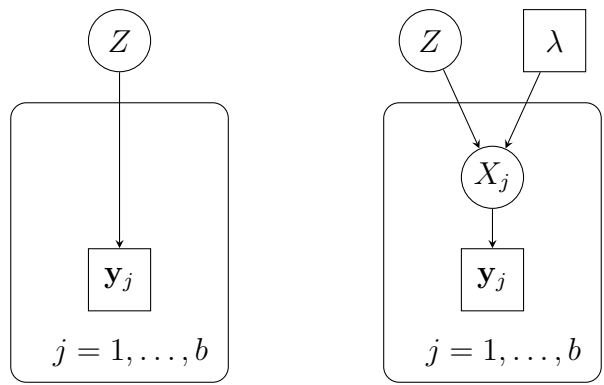

The instrumental hierarchical model is presented diagrammatically in Figure~\ref{fig:DAGs}.
The variables $X_{1:b}$ may be seen as `proxies' for $Z$ associated with each of the data subsets, which are conditionally independent given $Z$ and $\lambda$.
Loosely speaking, $\lambda$ represents the extent to which we allow the local variables $X_{1:b}$ to differ from the global variable $Z$.
In terms of computation, it is the separation of $Z$ from the subsets of the data $\mathbf{y}_{1:b}$, given $X_{1:b}$ introduced by the instrumental model, that can be exploited by distributed algorithms.

This approach to constructing an artificial joint target density is easily extended to accommodate random effects models, in which the original statistical model itself contains local variables associated with each data subset.
These variables may be retained in the resulting instrumental model, alongside the local proxies $X_{1:b}$ for $Z$.
A full description of the resulting model is presented in \citet{Rendell2019}.

The framework we describe is motivated by concepts in distributed optimisation, a connection that is also explored in the contemporaneous work of \citet{VonoEtAl2019}.
The global consensus optimisation problem \cite[see, e.g.,][Section~7]{BoydEtAl2011} is that of minimising a sum of functions on a common domain, under the constraint that their arguments are all equal to some global common value.
If for each $j \in \{1, \ldots, b\}$ one uses the Gaussian kernel density $K_{j}^{(\lambda)}(z,x) = \dnorm(x;z,\lambda)$, then taking the negative logarithm of \eqref{eq:jointZXdensity} gives
\begin{equation}
-\log \tilde{\pi}_{\lambda}(z,x_{1:b}) = 
C - \log\mu(z) - \sum_{j=1}^{b} \log f_{j}(x_{j}) + \frac{1}{2\lambda} \sum_{j=1}^{b} \left( z-x_{j} \right)^{2}
\label{eq:negativeLogarithm}
\end{equation}
where $C$ is a normalising constant.
Maximising $\pi(z)$ is equivalent to minimising this function under the constraint that $z=x_{j}$ for $j\in\{1,\dots,b\}$, which may be achieved using the alternating direction method of multipliers \citep{BertsekasTsitsiklis1989}. Specifically, \eqref{eq:negativeLogarithm} corresponds to the use of $1/\lambda$ as the penalty parameter in this procedure.

There are some similarities between this framework and Approximate Bayesian Computation \citep[ABC; see][for a review of such methods]{MarinEtAl2012}. In both cases one introduces a kernel that can be viewed as acting to smooth the likelihood.
In the case of \eqref{eq:jointZXdensity} the role of $\lambda$ is to control the scale of smoothing that occurs in the parameter space; the tolerance parameter used in ABC, in contrast, controls the extent of a comparable form of smoothing in the observation (or summary statistic) space.

\subsection{Distributed Metropolis-within-Gibbs}
\label{subsec:algorithm}

The instrumental model described forms the basis of our proposed global consensus framework; `global consensus Monte Carlo' is correspondingly the application of Monte Carlo methods to form an approximation of $\pi_{\lambda}$.
We focus here on the construction of a Metropolis-within-Gibbs Markov kernel that leaves $\tilde{\pi}_{\lambda}$ invariant.
If $\lambda$ is chosen to be sufficiently small, then the $Z$-marginal $\pi_{\lambda}$ provides an approximation of $\pi$.
Therefore given a chain with values denoted $(Z^{i},X_{1:b}^{i})$ for $i \in \{1,\ldots,N\}$, an empirical approximation of $\pi$ is given by
\begin{equation}
\pi_\lambda^N \coloneqq \frac{1}{N} \sum_{i=1}^{N} \delta_{Z^{i}},
\label{eq:gcmcEstimator}
\end{equation}
where $\delta_{z}$ denotes the Dirac measure at $z$.

The Metropolis-within-Gibbs kernel we consider utilises the full conditional densities
\begin{equation}
\tilde{\pi}_{\lambda}(x_{j} \mid z) \propto
K_{j}^{(\lambda)}(z,x_{j}) f_{j}(x_{j})
\label{eq:xFullConditional}
\end{equation}
for $j \in \{1,\ldots,b\}$, and
\begin{equation}
\tilde{\pi}_{\lambda}(z\mid x_{1:b}) \propto
\mu(z) \prod_{j=1}^{b} K_{j}^{(\lambda)}(z,x_{j}),
\label{eq:zFullConditional}
\end{equation}
where \eqref{eq:xFullConditional} follows from the mutual conditional independence of $X_{1:b}$ given $Z$.
Here we observe that $K_{j}^{(\lambda)}(z,x_{j})$ simultaneously provides a pseudo-prior for $X_{j}$ and a pseudo-likelihood for $Z$.

We define $M_{1}^{(\lambda)}$ to be a $\tilde{\pi}_{\lambda}$-invariant Markov kernel that fixes $z$; we may write
\begin{equation}
M_{1}^{(\lambda)}((z,x_{1:b}); \diff (z',x{}_{1:b}')) = 
\delta_{z}(\diff z') \prod_{j=1}^{b} P_{j,z}^{(\lambda)}(x_{j},\diff x_{j}'),
\label{eq:M1kernel}
\end{equation}
where for each $j$, $P_{j,z}^{(\lambda)}(x_{j},\cdot)$ is a Markov kernel leaving \eqref{eq:xFullConditional} invariant.
We similarly define $M_{2}^{(\lambda)}$ to be a $\tilde{\pi}_{\lambda}$-invariant Markov kernel that fixes $x_{1:b}$,
\begin{equation}
M_{2}^{(\lambda)}((z,x_{1:b}); \diff (z',x{}_{1:b}')) =
\left[ \prod_{j=1}^{b} \delta_{x_{j}}(\diff x{}_{j}') \right] P^{(\lambda)}_{x_{1:b}}(z,\diff z'),
\label{eq:M2kernel}
\end{equation}
where $P_{x_{1:b}}^{(\lambda)}(z, \cdot)$ is a Markov kernel leaving \eqref{eq:zFullConditional} invariant.

Using these Markov kernels we construct an MCMC kernel that leaves $\tilde{\pi}_{\lambda}$ invariant; we present the resulting sampling procedure as Algorithm~\ref{alg:gcmcMCMC}.
In the special case in which one may sample exactly from the conditional distributions \eqref{eq:xFullConditional}--\eqref{eq:zFullConditional}, this algorithm takes the form of a Gibbs sampler.

\begin{algorithm}
	\caption{Global consensus Monte Carlo: MCMC algorithm}
	\label{alg:gcmcMCMC}
	Fix $\lambda > 0$. Set initial state $(Z^{0}, X_{1:b}^{0})$; choose chain length $N$.

	\medskip

	For $i = 1, \dots, N$:

	\begin{itemize}
	\item For $j \in \{1, \dots, b\}$, sample $X_{j}^{i} \sim P_{j, Z^{i-1}}^{(\lambda)}(X_{j}^{i-1}, \cdot)$.
	\item Sample $Z^{i} \sim P_{X_{1:b}^{i}}^{(\lambda)}(Z^{i-1}, \cdot)$.
    \end{itemize}

	Return $(Z^{i}, X_{1:b}^{i})_{i=1}^N$.
\end{algorithm}

The interest from a distributed perspective is that the full conditional density \eqref{eq:xFullConditional} of each $X_{j}$, for given values $x_{j}$ and $z$, depends only on the $j$th block of data (through the partial likelihood $f_{j}$) and may be computed on the $j$th machine.
Within Algorithm~\ref{alg:gcmcMCMC}, the sampling of each $X_{j}^{i}$ from $P_{j,Z^{i-1}}^{(\lambda)}(X_{j}^{i-1},\cdot)$ may therefore occur on the $j$th machine; these $X_{1:b}^{i}$ may then be communicated to a central machine that draws $Z^{i}$.

Our approach has particular benefits when sampling exactly from \eqref{eq:xFullConditional} is not possible, in which case Algorithm~\ref{alg:gcmcMCMC} takes a Metropolis-within-Gibbs form.
One may choose the Markov kernels $P_{j,z}^{(\lambda)}$ to comprise multiple iterations of an MCMC kernel leaving \eqref{eq:xFullConditional} invariant; indeed multiple computations of each $f_j$ (and therefore multiple accept/reject steps) may be conducted on each of the $b$ nodes, without requiring communication between machines.
This stands in contrast to more straightforward MCMC approaches directly targeting $\pi$, in which such communication is required for each evaluation of \eqref{eq:piOriginal}, and therefore for every accept/reject step. 
Similar to such a `direct' MCMC approach, each iteration of Algorithm~\ref{alg:gcmcMCMC} requires communication to and from each of the $b$ machines on which the data are stored; but in cases where the communication latency is high, the resulting sampler will spend a greater proportion of time exploring the state space.
This may in turn result in faster mixing (e.g.\ with respect to wall-clock time).
We further discuss this comparison, and the role of communication latency, in Section~\ref{subsec:repeatedMCMCkernels} of the supplement.

The setting in which Algorithm~\ref{alg:gcmcMCMC} describes a Gibbs sampler, with all variables drawn exactly from their full conditional distributions, is particularly amenable to analysis.
We provide such a study in Section~\ref{sec:Gaussians} of the supplement.
This analysis may also be informative about the more general Metropolis-within-Gibbs setting, when $P_{j,z}^{(\lambda)}$ comprises enough MCMC iterations to exhibit good mixing.

\subsection{Choosing the regularisation parameter}
\label{subsec:choosingLambda}
 
For $\varphi : \sfE \to \reals$ we may estimate $\pi(\varphi)$ using \eqref{eq:gcmcEstimator} as $\pi_\lambda^N(\varphi)$. The regularisation parameter $\lambda$ here takes the role of a tuning parameter; we can view its effect on the mean squared error of such estimates using the bias--variance decomposition,
\begin{equation}
\label{eq:gcmc_mse}
\E \! \left[\left( \pi_\lambda^N(\varphi) - \pi(\varphi) \right)^2\right] =
[\pi_\lambda(\varphi) - \pi(\varphi)]^2
+ \var[\pi_\lambda^N(\varphi)],
\end{equation}
which holds exactly when $\E[\pi_\lambda^N(\varphi)] = \pi_\lambda(\varphi)$.
In many practical cases this decomposition will provide a very accurate approximation for large $N$, as the squared bias of $\pi_\lambda^N(\varphi)$ is typically asymptotically negligible in comparison to its variance.

The decomposition \eqref{eq:gcmc_mse} separates the contributions to the error from the bias introduced by the instrumental model and the variance associated with the MCMC approximation.
If $\lambda$ is too large, the squared bias term in \eqref{eq:gcmc_mse} can dominate while if $\lambda$ is too small, the Markov chain may exhibit poor mixing due to strong conditional dependencies between $X_{1:b}$ and $Z$, and so the variance term in \eqref{eq:gcmc_mse} can dominate.

It follows that $\lambda$ should ideally be chosen in order to balance these two considerations.
We provide a theoretical analysis of the role of $\lambda$ in Section~\ref{sec:Gaussians} of the supplement, by considering a simple Gaussian setting.
In particular we show that for a fixed number of data, one should scale $\lambda$ with the number of samples $N$ as $\mathcal{O}(N^{-1/3})$ in order to minimise the mean squared error.
We also consider the consistency of the approximate posterior $\pi_{\lambda}$ as the number of data $n$ tends to infinity.
Specifically, suppose these are split equally into $b$ blocks; considering $\lambda$ and $b$ as functions of $n$, we find that $\pi_{\lambda}$ exhibits posterior consistency if $\lambda/b$ decreases to $0$ as $n \to \infty$, and that credible intervals have asymptotically correct coverage if $\lambda/b$ decreases at a rate strictly faster than $n^{-1}$.

As an alternative to selecting a single value of $\lambda$, we propose in Section~\ref{sec:SMCimplementation} a Sequential Monte Carlo sampler employing Markov kernels formed via Algorithm~\ref{alg:gcmcMCMC}. In this manner a decreasing sequence of $\lambda$ values may be considered, which may result in lower-variance estimates for small $\lambda$ values; we also describe a possible bias correction technique.

\subsection{Related approaches}
\label{subsec:relatedApproaches}

As previously mentioned, a Gibbs sampler construction essentially corresponding to Algorithm~\ref{alg:gcmcMCMC} has independently been proposed by \citet{VonoEtAl2019}.
Their main objective is to improve algorithmic performance when computation of the full posterior density is intensive by constructing full conditional distributions that are more computationally tractable, for which they primarily consider a setting in which $b=1$.
In contrast, we focus on the exploitation of this framework in distributed settings (i.e.\ with $b>1$), in the manner described in Section~\ref{subsec:algorithm}.

The objectives of our algorithm are similar, but not identical, to those of the previously-introduced `embarrassingly parallel' approaches proposed by many authors. 
These reduce the costs of communication latency to a near-minimum by simulating a separate MCMC chain on each machine; typically, the chain on the $j$th machine is invariant with respect to a `subposterior' distribution with density proportional to $\mu(z)^{1/b} f_{j}(z)$.
Communication is necessary only for the final aggregation step, in which an approximation of the full posterior is obtained using the samples from all $b$ chains.

A well-studied approach within this framework is Consensus Monte Carlo \citep{ScottEtAl2016}, in which the post-processing step amounts to forming a `consensus chain' by weighted averaging of the separate chains.
In the case that each subposterior density is Gaussian this approach can be used to produce samples asymptotically distributed according to $\pi$, by weighting each chain using the precision matrices of the subposterior distributions.
Motivated by Bayesian asymptotics, the authors suggest using this approach more generally.
In cases where the subposterior distributions exhibit near-Gaussianity this performs well, with the final `consensus chain' providing a good approximation of posterior expectations.
However, there are no theoretical guarantees associated with this approach in settings in which the subposterior densities are poorly approximated by Gaussians.
In such cases consensus Monte Carlo sometimes performs poorly in forming an approximation of the posterior $\pi$ \citep[as in examples of][]{WangEtAl2015,SrivastavaEtAl2015,DaiEtAl2019}, and so the resulting estimates of integrals $\pi(\varphi)$ exhibit high bias.

Various authors \citep[e.g.][]{MinskerEtAl2014,RabinovichEtAl2015,SrivastavaEtAl2015,WangEtAl2015} have therefore proposed alternative techniques for utilising the values from each of these chains in order to approximate posterior expectations, each of which presents benefits and drawbacks.
For example, \citet{NeiswangerEtAl2014} propose a strategy based on kernel density estimation; based on this approach, \citet{Scott2017} suggests a strategy based on finite mixture models, though notes that both methods may be impractical in high-dimensional settings.

An aggregation procedure proposed by \citet{WangDunson2013} bears some relation to our proposed framework, being based on the application of Weierstrass transforms to each subposterior density.
The resulting smoothed densities are analogous to \eqref{eq:smoothedLikelihood}, which represents a smoothed form of the partial likelihood $f_j$.
As well as proposing an aggregation method based on rejection sampling, the authors propose a technique for `refining' an initial posterior approximation, which may be expressed in terms of a Gibbs kernel on an extended state space.
Comparing with our framework, this is analogous to applying Algorithm~\ref{alg:gcmcMCMC} for one iteration with $N$ different initial values.

A potential issue common to these approaches is the treatment of the prior density $\mu$. 
Each subposterior density is typically assigned an equal share of the prior information in the form of a fractionated prior density $\mu(z)^{1/b}$, but it is not clear when this approach is satisfactory.
For example, suppose $\mu$ belongs to an exponential family; any property that is not invariant to multiplying the canonical parameters by a constant will not be preserved in the fractionated prior.
As such if $\mu(z)^{1/b}$ is proportional to a valid probability density function of $z$, then the corresponding distribution may be qualitatively very different to the full prior.
Although \citet{ScottEtAl2016} note that fractionated priors perform poorly on some examples (for which tailored solutions are provided), no other general way of assigning prior information to each block naturally presents itself.
In contrast our approach avoids this problem entirely, with $\mu$ providing prior information for $Z$ at the `global' level.

Finally, we believe this work is complementary to other distributed algorithms lying outside of the `embarrassingly parallel' framework \citep[including][]{XuEtAl2014,JordanEtAl2019}, and to approaches that aim to reduce the amount of computation associated with each likelihood calculation on a single node, e.g.\ by using only a subsample or batch of the data \citep[as in][]{KorattikaraEtAl2014, BardenetEtAl2014, HugginsEtAl2016, MaclaurinAdams2014}.

\section{Sequential Monte Carlo approach}
\label{sec:SMCimplementation}

As discussed in Section~\ref{subsec:choosingLambda}, as $\lambda$ approaches $0$ estimators $\pi_{\lambda}^{N}(\varphi)$ formed using \eqref{eq:gcmcEstimator} exhibit lower bias but higher variance, due to poorer mixing of the resulting Markov chain.
In order to obtain lower-variance estimators for $\lambda$ values close to $0$, we consider the use of Sequential Monte Carlo (SMC) methodology to generate suitable estimates for a sequence of $\lambda$ values.

SMC methodology employs sequential importance sampling and resampling; recent surveys include \citet{DoucetJohansen2011} and \citet{DoucetLee2018}.
We consider here approximations of a sequence of distributions with densities $\tilde{\pi}_{\lambda_0}, \tilde{\pi}_{\lambda_1}, \ldots$, where $\lambda_0,\ldots,\lambda_n$ is a decreasing sequence.
The procedure we propose, specified in Algorithm~\ref{alg:gcmcSMC}, is an SMC sampler within the framework of \citet{DelMoralEtAl2006}.
This algorithmic form was first proposed by \citet{GilksBerzuini2001} and \citet{Chopin2002} in different settings, building upon ideas in \citet{Crooks1998}, \citet{Neal2001}.

\begin{algorithm}
	\caption{Global consensus Monte Carlo: SMC algorithm}
	\label{alg:gcmcSMC}
	Fix a decreasing sequence $(\lambda_{0}, \lambda_{1}, \dots, \lambda_{n})$. Set number of particles $N$.

	\medskip

	Initialise:

	\begin{itemize}
		\item For $i \in \{1, \dots, N\}$, sample $\zeta_{0}^{i} = (Z_{0}^{i},X_{0,1:b}^{i}) \sim \tilde{\pi}_{\lambda_{0}}$ and set $W_{0}^{i} \gets 1$.
	\end{itemize}

	For $p = 1, \dots, n$:

	\begin{enumerate}
		\item For $i \in \{1,\ldots,N\}$, set $\tilde{W}_{p}^{i} \gets W_{p-1}^{i} w_{p}(\zeta_{p-1}^{i})$, where
    \[
    w_{p}(z,x_{1:b}) \coloneqq \frac{\tilde{\pi}_{\lambda_p}(z,x_{1:b})}{\tilde{\pi}_{\lambda_{p-1}}(z,x_{1:b})} = \prod_{j=1}^{b} \frac{K_{j}^{(\lambda_{p})}(z,x_{j})}{K_{j}^{(\lambda_{p-1})}(z,x_{j})}.
    \]
		\item Optionally, carry out a resampling step: for $i \in \{1,\ldots,N\}$,
    \begin{itemize}
    \item Sample $A_{p-1}^{i} \sim {\rm Categorical}(\tilde{W}_p^1,\ldots,\tilde{W}_p^N)$ independently.
    \item Set $W_{p}^{i} \gets 1$.
    \end{itemize}
    Otherwise: for $i \in \{1,\ldots,N\}$ set $A_{p-1}^{i} \gets i$, $W_{p}^{i} \gets \tilde{W}_{p}^{i}$.
		\item For $i \in \{1,\ldots,N\}$, sample $\zeta_{p}^{i} = (Z_{p}^{i},X_{p,1:b}^{i}) \sim M_p(\zeta_{p-1}^{A_{p-1}^{i}}, \cdot)$, where $M_p$ is a $\tilde{\pi}_{\lambda_{p}}$-invariant MCMC kernel constructed in the manner of Algorithm~\ref{alg:gcmcMCMC}.
    \item Optionally, store the particle approximation of $\tilde{\pi}_{\lambda_p}$,
    \[
    \tilde{\pi}^N_{\lambda_p} \coloneqq \frac{\sum_{i=1}^N W_p^i \delta_{\zeta_p^i}}{\sum_{i=1}^N W_p^i}.
    \]
	\end{enumerate}

\end{algorithm}

The algorithm presented involves simulating particles using $\tilde{\pi}_\lambda$-invariant Markov kernels, and has a genealogical structure imposed by the ancestor indices $A_p^i$ for $p \in \{0,\ldots,n-1\}$ and $i \in \{1,\ldots,N\}$.
The specific scheme for simulating the ancestor indices here is known as multinomial resampling; other schemes can be used \citep[see][for a summary of some schemes and their properties]{DoucEtAl2005,GerberEtAl2019}.
We use this simple scheme here as it validates the use of the variance estimators used in Section~\ref{subsec:biasCorrection}.
This optional resampling step is used to prevent the degeneracy of the particle set; a common approach is to carry out this step whenever the particles' effective sample size \citep{LiuChen1995} falls below a pre-determined threshold.

Under weak conditions $\tilde{\pi}^N_{\lambda_{p}}(\varphi)$ converges almost surely to $\tilde{\pi}_{\lambda_{p}}(\varphi)$ as $N \to \infty$.
One can also define the particle approximations of $\pi_{\lambda_{p}}$ via
\begin{equation}
{\pi}^N_{\lambda_{p}} \coloneqq \frac{\sum_{i=1}^N W_{p}^{i} \delta_{Z_{p}^{i}}}{\sum_{i=1}^N W_{p}^{i}},
\label{eq:SMCapproximation_marginal}
\end{equation}
where $Z_{p}^{i}$ is the first component of the particle $\zeta_{p}^{i}$.

Although the algorithm is specified for simplicity in terms of a fixed sequence $\lambda_0,\ldots,\lambda_n$, a primary motivation for the SMC approach is that the sequence used can be determined adaptively while running the algorithm.
A number of such procedures have been proposed in the literature in the context of tempering, allowing each value $\lambda_p$ to be selected based on the particle approximation of $\tilde{\pi}_{\lambda_{p-1}}$.
For example, \citet{JasraEtAl2011} propose a procedure that controls the decay of the particles' effective sample size.
Within the examples in Section~\ref{sec:examples} we employ a procedure proposed by \citet{ZhouEtAl2016}, which generalises this approach to settings in which resampling is not conducted in every iteration, aiming to control directly the dissimilarity between successive distributions.
A possible approach to determining when to terminate the procedure, based on minimising the mean squared error \eqref{eq:gcmc_mse}, is detailed in Section~\ref{subsec:stoppingRule} of the supplement.

With regard to initialisation, if it is not possible to sample from  $\tilde{\pi}_{\lambda_{0}}$ one could instead use samples obtained by importance sampling, or one could initialise an SMC sampler with some tractable distribution and use tempering or similar techniques to reach $\tilde{\pi}_{\lambda_{0}}$.
At the expense of the introduction of an additional approximation, an alternative would be to run a $\tilde{\pi}_{\lambda_{0}}$-invariant Markov chain, and obtain an initial collection of particles by thinning the output \citep[an approach that may be validated using results of][]{FinkeEtAl2018}.
Specifically, one could use Algorithm~\ref{alg:gcmcMCMC} to generate such samples for some large $\lambda_{0}$, benefiting from its good mixing and low autocorrelation when $\lambda$ is sufficiently large.
The effect of Algorithm~\ref{alg:gcmcSMC} may then be seen as refining or improving the resulting estimators, by bringing the parameter $\lambda$ closer to zero.

Other points in favour of this approach are that many of the particle approximations \eqref{eq:SMCapproximation_marginal} can be used to form a final estimate of $\pi(\varphi)$ as explored in Section~\ref{subsec:biasCorrection}, and that SMC methods can be more robust to multimodality of $\pi$ than simple Markov chain schemes.
We finally note that in such an SMC sampler, a careful implementation of the MCMC kernels used may allow the inter-node communication to be interleaved with likelihood computations associated with the particles, thereby reducing the costs associated with communication latency.

\subsection{Bias correction using local linear regression}
\label{subsec:biasCorrection}

We present an approach to use many of the particle approximations produced by Algorithm~\ref{alg:gcmcSMC}. A natural idea is to regress the values of $\pi_\lambda^N(\varphi)$ on $\lambda$, extrapolating to $\lambda = 0$ to obtain an estimate of $\pi(\varphi)$.
A similar idea has been used for bias correction in the context of Approximate Bayesian Computation, albeit not in an SMC setting, regressing on the discrepancy between the observed data and simulated pseudo-observations \citep{BeaumontEtAl2002, BlumFrancois2010}.

Under very mild assumptions on the transition densities $K_{j}^{(\lambda)}$, $\pi_\lambda(\varphi)$ is smooth as a function of $\lambda$.
Considering a first-order Taylor expansion of this function, a simple approach is to model the dependence of $\pi_\lambda(\varphi)$ on $\lambda$ as linear, for $\lambda$ sufficiently close to $0$.
Having determined a subset of the values of $\lambda$ used for which a linear approximation is appropriate (we propose a heuristic approach in Section~\ref{subsec:inclusionProcedure} of the supplement) one can use linear least squares to carry out the regression.
To account for the SMC estimates $\pi_{\lambda_{p}}^N(\varphi)$ having different variances, we propose the use of weighted least squares, with the `observations' $\pi_{\lambda_{p}}^N(\varphi)$ assigned weights inversely proportional to their estimated variances; we describe methods for computing such variance estimates in Section~\ref{subsec:varianceEstimation}.
A bias-corrected estimate of $\pi(\varphi)$ is then obtained by extrapolating the resulting fit to $\lambda = 0$, which corresponds to taking the estimated intercept term.

To make this explicit, first consider the case in which $\varphi: \sfE \to \reals$, so that the estimates $\pi_\lambda^N(\varphi)$ are univariate. For each value $\lambda_{p}$ denote the corresponding SMC estimate by $\eta_{p} \coloneqq \pi_{\lambda_{p}}^N(\varphi)$, and let $v_{p}$ denote some proxy for the variance of this estimate.
Then for some set of indices $S \coloneqq \{p^*, \dots, n\}$ chosen such that the relationship between $\eta_{p}$ and $\lambda_{p}$ is approximately linear for $p \in S$, a bias-corrected estimate for $\pi(\varphi)$ may be computed as
\begin{equation}
\pi^{\mathrm{BC}}_{S}(\varphi) \coloneqq
\tilde{\eta}_{S} -
\tilde{\lambda}_{S} \frac{\sum_{p \in S} (\lambda_{p}-\tilde{\lambda}_{S}) (\eta_{p}-\tilde{\eta}_{S}) / v_{p}}
{\sum_{p \in S} (\lambda_{p}-\tilde{\lambda}_{S})^{2} / v_{p}},
\label{eq:biasCorrectedEstimate}
\end{equation}
where $\tilde{\lambda}_{S}$ and $\tilde{\eta}_{S}$ denote weighted means given by
\[
\tilde{\lambda}_{S} \coloneqq
\frac{\sum_{p \in S}\lambda_{p}/v_{p}}{\sum_{p \in S} 1/v_{p}},
\qquad
\tilde{\eta}_{S} \coloneqq
\frac{\sum_{p \in S}\eta_{p}/v_{p}}{\sum_{p \in S} 1/v_{p}}.
\]
The formal justification of this estimate assumes that the observations are uncorrelated, which does not hold here.
We demonstrate in Section~\ref{sec:examples} examples on which this simple approach is nevertheless effective, but in principle one could use generalised least squares combined with some approximation of the full covariance matrix of the SMC estimates.

In the more general case where $\varphi: \sfE \to \reals^d$ for $d>1$, we propose simply evaluating \eqref{eq:biasCorrectedEstimate} for each component of this quantity separately, which corresponds to fitting an independent weighted least squares regression to each component. This facilitates the use of the variance estimators described in the following section, though in principle one could use multivariate weighted least squares or other approaches.

\subsection{Variance estimation for weighted least squares}
\label{subsec:varianceEstimation}

We propose the weighted form of least squares here since, as the values of $\lambda$ used in the SMC procedure approach zero, the estimators generated may increase in variance: partly due to poorer mixing of the MCMC kernels as previously described, but also due to the gradual degeneracy of the particle set.
In order to estimate the variances of estimates generated using SMC, several recent approaches have been proposed that allow this estimation to be carried out online by considering the genealogy of the particles. Using any such procedure, one may estimate the variance of $\pi_\lambda^N(\varphi)$ for each $\lambda$ value considered by Algorithm~\ref{alg:gcmcSMC}, with these values used for each $v_{p}$ in \eqref{eq:biasCorrectedEstimate}.

Within our examples, we use the estimator proposed by \citet{LeeWhiteley2018}, which for fixed $N$ coincides with an earlier proposal of \citet{ChanEtAl2013} (up to a multiplicative constant).
Specifically, after each step of the SMC sampler we compute an estimate of the \emph{asymptotic} variance of each estimate $\pi_{\lambda_{p}}^{N}(\varphi)$; that is, the limit of $N \var [\pi_{\lambda_{p}}^{N}(\varphi)]$  as $N \to \infty$.
While this is not equivalent to computing the true variance of each estimate, for fixed large $N$ the relative sizes of these asymptotic variance estimates should provide a useful indicator of the relative variances of each estimate $\pi_{\lambda}^{N}(\varphi)$.
In Section~\ref{subsec:gaussianSMCexample} of the supplement we show empirically that inversely weighting the SMC estimates according to these estimated variances can result in more stable bias-corrected estimates as the particle set degenerates.
We also explain in Section~\ref{subsec:stoppingRule} how these estimated variances can be used within a rule to determine when to terminate the algorithm.

The asymptotic variance estimator described is consistent in $N$.
However, if in practice resampling at the $p$th time step causes the particle set to degenerate to having a single common ancestor, then the estimator evaluates to zero, and so it is impossible to use this value as the inverse weight $v_p$ in \eqref{eq:biasCorrectedEstimate}.
Such an outcome may be interpreted as a warning that too few particles have been used for the resulting SMC estimates to be reliable, and that a greater number should be used when re-running the procedure.

\section{Examples}
\label{sec:examples}

\subsection{Log-normal toy example}
\label{subsec:lognormalExample}

To compare the posterior approximations formed by our global consensus framework with those formed by some embarrassingly parallel approaches discussed in Section~\ref{subsec:relatedApproaches}, we conduct a simulation study based on a simple model.
Let $\dlnorm(x;\mu,\sigma^{2})$ denote the density of a log-normal distribution with parameters $(\mu, \sigma^{2})$; that is,
\[
\dlnorm(x;\mu,\sigma^{2})=\frac{1}{x\sqrt{2\pi\sigma^{2}}}\exp\left(-\frac{(\log(x)-\mu)^{2}}{2\sigma^{2}}\right).
\]
One may consider a model with prior density $\mu(z)=\dlnorm(z;\mu_{0},\sigma_{0}^{2})$ and likelihood contributions $f_{j}(z)=\dlnorm(\log(\mu_{j});\log(z),\sigma_{j}^{2})$ for $j \in \{1,\dots,b\}$.
This may be seen as a reparametrisation of the Gaussian model in Section~\ref{sec:Gaussians} of the supplement, in which each likelihood contribution is that of a data subset with a Gaussian likelihood.
This convenient setting allows for the target distribution $\pi$ to be expressed analytically.
For the implementation of the global consensus algorithm, we choose Markov transition kernels given by $K_{j}^{(\lambda)}(z,x)=\dlnorm(x;\log(z),\lambda)$ for each $j \in \{1, \dots, b\}$, which satisfy Assumption~\ref{ass:lambda}; this allows for exact sampling from all the full conditional distributions.

As a toy example to illustrate the effects of non-Gaussian partial likelihoods we consider a case in which $f_{j}(z)=\dlnorm(\log(\mu_{j});\log(z),1)$ for each $j$, and $\mu(z)=\dlnorm(z;0,25)$.
Here we took $b=32$, and selected the location parameters $\mu_{j}$ as i.i.d.\ samples from a standard normal distribution. 
We ran Global Consensus Monte Carlo (GCMC) using the Gibbs sampler form of Algorithm~\ref{alg:gcmcMCMC}, for values of $\lambda$ between $10^{-5}$ and $10$.
For comparison we also drew samples from each subposterior distribution as defined in Section~\ref{subsec:relatedApproaches}, combining the samples using various approaches.
These are the Consensus Monte Carlo (CMC) averaging of \citet{ScottEtAl2016}; the nonparametric density product estimation (NDPE) approach of \citet{NeiswangerEtAl2014}; and the Weierstrass rejection sampling (WRS) combination technique of \citet{WangDunson2013}, using their R implementation (\url{https://github.com/wwrechard/weierstrass}).
In each case we ran the algorithm $25$ times, drawing $N = 10^{5}$ samples.

\begin{table}
\begin{centering}
\begin{tabular}{llccc}
\multicolumn{2}{l}{}
 & $\varphi(z) = z$
 & $\varphi(z) = z^5$
 & $\varphi(z) = \log(z)$ \\
\multicolumn{2}{l}{Algorithm}
& ($\pi(\varphi) = 1.141$)
& ($\pi(\varphi) = 2.644$)
& ($\pi(\varphi) = 0.1164$) \\
\hline
\multirow[t]{7}{*}{GCMC} & $\lambda=10^{1}$
 & $1.329 \pm 0.003$
 & $121.154 \pm 10.487$
 & $0.1151 \pm 0.0019$ \\
 & $\lambda=10^{0}$
 & $1.159 \pm 0.002$
 & $\hphantom{00}3.901 \pm \hphantom{0}0.037$
 & $\mathbf{0.1165 \pm 0.0014}$ \\
 & $\lambda=10^{-1}$
 & $\mathbf{1.144 \pm 0.003}$
 & $\mathbf{\hphantom{00}2.763 \pm \hphantom{0}0.044}$
 & $0.1173 \pm 0.0030$ \\
 & $\lambda=10^{-2}$
 & $1.140 \pm 0.011$
 & $\hphantom{00}2.648 \pm \hphantom{0}0.143$
 & $0.1150 \pm 0.0090$ \\
 & $\lambda=10^{-3}$
 & $1.142 \pm 0.022$
 & $\hphantom{00}2.661 \pm \hphantom{0}0.295$
 & $0.1191 \pm 0.0199$ \\
 & $\lambda=10^{-4}$
 & $1.120 \pm 0.077$
 & $\hphantom{00}3.505 \pm \hphantom{0}1.136$
 & $0.1630 \pm 0.0651$ \\
 & $\lambda=10^{-5}$
 & $1.400 \pm 0.110$
 & $\hphantom{00}6.195 \pm \hphantom{0}2.217$
 & $0.3283 \pm 0.0810$ \\
 \hline
\multicolumn{2}{l}{CMC}
 & $1.073 \pm 0.010$
 & $\hphantom{0}16.092 \pm \hphantom{0}5.675$
 & $0.0135 \pm 0.0095$ \\
\multicolumn{2}{l}{NDPE}
 & $1.148 \pm 0.029$
 & $\hphantom{00}2.800 \pm \hphantom{0}0.385$
 & $0.1231 \pm 0.0246$ \\
\multicolumn{2}{l}{WRS}
 & $1.111 \pm 0.007$
 & $\hphantom{00}2.444 \pm \hphantom{0}0.086$
 & $0.0862 \pm 0.0063$ \\
\hline
\end{tabular}
\par\end{centering}
\caption{
	True values and estimates of $\pi(\varphi)$, for various test functions $\varphi$, for the log-normal toy model.
	Estimates obtained using Global Consensus Monte Carlo with various values of $\lambda$, and three embarrassingly parallel methods (see main text for abbreviations).
    For each method the mean estimate $\pm$ Monte Carlo standard error is presented, as computed over $25$ replicates; the estimator corresponding to the lowest mean squared error is printed in bold.
}
\label{tab:lognormalResults1}
\end{table}

To demonstrate the role of $\lambda$ in the bias--variance decomposition \eqref{eq:gcmc_mse}, Table~\ref{tab:lognormalResults1} presents the means and standard deviations of estimates of $\pi(\varphi)$, for various test functions $\varphi$.
In estimating the first moment of $\pi$, GCMC generates a low-bias estimator when $\lambda$ is chosen to be sufficiently small; however, as expected, the variance of such estimators increases when very small values of $\lambda$ are chosen.
While the other methods produce estimators of reasonably low variance, these exhibit somewhat higher bias.
For CMC the bias is especially pronounced when estimating higher moments of the posterior distribution, as exemplified by the estimates of the fifth moment.
Note however that high biases are also introduced when using Global Consensus Monte Carlo with large values of $\lambda$ (as seen here with $\lambda = 10$), for which $\pi_{\lambda}$ is a poor approximation of $\pi$.

Also of note are estimates of $\int \log(z) \pi(z) \diff z$, corresponding to the mean of the Gaussian model of which this a reparametrisation.
While Global Consensus Monte Carlo performs well across a range of $\lambda$ values, the other methods perform less favourably; Consensus Monte Carlo produces an estimate that is incorrect by an order of magnitude.
While this could be solved by a simple reparametrisation of the problem in this case, in more general settings no such straightforward solution may exist.

In Section~\ref{subsec:additionalLognormalExample} of the supplement we present second example based on a log-normal model, demonstrating the robustness of these methods to permutation and repartitioning of the data.

\subsection{Logistic regression}
\label{subsec:logRegExample}

Binary logistic regression models are commonly used in settings related to marketing.
In web design for example, A/B testing may be used to determine which content choices lead to maximised user interaction, such as the user clicking on a product for sale.

We assume that we have a data set of size $n$ formed of responses $\eta_{\ell}\in\{-1,1\}$, and vectors $\xi_{\ell}\in\{0,1\}^{d}$ of binary covariates, where $\ell \in \{1,\dots,n\}$.
The likelihood contribution of each block of data then takes the form $f_{j}(z)=\prod_{\ell} S(\eta_{\ell}z^{\mathsf{T}} \xi_{\ell})$, $z\in\reals^{d}$, where the product is taken over those indices $\ell$ included in the $j$th block of data, and $S$ denotes the logistic function, $S(x)=(1+\exp(x))^{-1}$.

For the prior $\mu$, we use a product of independent zero-mean Gaussians, with standard deviation $20$ for the parameter corresponding to the intercept term, and $5$ for all other parameters.
For the Markov transition densities in GCMC, we use multivariate Gaussian densities: $K_{j}^{(\lambda)}(z,x)=\dnorm(x;z,\lambda I)$ for each $j \in \{1, \dots\ b\}$.

We investigated several such simulated data sets and the efficacy of various approaches in approximating the true posterior $\pi$.
To illustrate the bias--variance trade-off described in Section~\ref{subsec:choosingLambda}, in the presentation of these results we focus on the estimation of the posterior first moment; denoting the identity function on $\reals^{d}$ by $\Id$, we may write this as $\pi(\Id)$.
While our global consensus approach was consistently successful in forming estimators with low mean squared error in each component, in low-dimensional settings the application of Consensus Monte Carlo often resulted in marginal improvements.
However, in many higher-dimensional settings, the estimators resulting from CMC exhibited relatively large biases. 

We present here an example in which the $d$ predictors correspond to $p$ binary input variables, their pairwise products, and an intercept term, so that $d = 1 + p + \binom{p}{2}$. 
In settings where the interaction effects corresponding to these pairwise products are of interest, the dimensionality $d$ of the space can be very large compared to $p$.
We used a simulated data set with $p = 20$ input variables, resulting in a parameter space of dimension $d = 211$.
The data comprise $n=80{,}000$ observations, split into $b=8$ equally-sized blocks.
Each observation of the $20$ binary variables was generated from a Bernoulli distribution with parameter $0.1$, and for each vector of covariates, the response was generated from the correct model, for a fixed underlying parameter vector $z^*$.

\subsubsection{Metropolis-within-Gibbs}

We applied GCMC for values of $\lambda$ between $10^{-4}$ and $1$.
We used a Metropolis-within-Gibbs formulation of Algorithm~\ref{alg:gcmcMCMC}, sampling directly from the Gaussian conditional distribution of $Z$ given $X_{1:b}$.
To sample approximately from the conditional distributions of each $X_j$ given $Z$ we used Markov kernels $P_{j,z}^{(\lambda)}$ comprising $k = 20$ iterations of a random walk Metropolis kernel.

As mentioned in Section~\ref{subsec:algorithm}, in settings of high communication latency our approach allows a greater proportion of wall-clock time to be spent on likelihood contributions, compared to an MCMC chain directly targeting the full posterior $\pi$.
To compare across settings, we therefore consider an abstracted distributed setting as discussed in Section~\ref{subsec:repeatedMCMCkernels} of the supplement, here assuming that the latency is $10$ times the time taken to compute each partial likelihood $f_{j}$ (in the notation of Section~\ref{subsec:repeatedMCMCkernels}, $C = 10\ell$).

We also compare with the same embarrassingly parallel approaches as in Section~\ref{subsec:lognormalExample} (CMC, NDPE, WRS), which are comparatively unaffected by communication latency. 
For these methods, we again used random walk Metropolis to draw samples from each subposterior distribution.
To ease computation, we thinned these chains before applying the combination step; in practice, the estimators obtained using these thinned chains behaved very similarly to those obtained using all subposterior samples.

To provide a `ground truth' against which to compare the results we ran a random walk Metropolis chain of length $500{,}000$ targeting $\pi$.
For all our random walk Metropolis samplers we used Gaussian proposal kernels.
To determine the covariance matrices of these, we formed a Laplace approximation of the target density following the approach of \citet{ChopinEtAl2017}, scaling the resulting covariance matrix optimally according to results of \citet{RobertsRosenthal2001}.

\begin{table}
\begin{centering}
\begin{tabular}{llc}
\multicolumn{2}{l}{Algorithm}
 & Mean sum of squared errors
 \\
\hline
\multirow[t]{5}{*}{GCMC} & $\lambda=10^{0}$
 & $0.1835$
 \\
 & $\lambda=10^{-1}$
 & $0.1379$
 \\
 & $\lambda=10^{-2}$
 & $0.0770$
 \\
 & $\lambda=10^{-3}$
 & $\mathbf{0.0478}$
 \\
 & $\lambda=10^{-4}$
 & $0.0662$
 \\
 \multicolumn{2}{l}{CMC}
 & $0.3710$
 \\
 \multicolumn{2}{l}{NDPE}
 & $0.8476$
 \\
 \multicolumn{2}{l}{WRS}
 & $0.6402$
 \\
 \multicolumn{2}{l}{Direct MCMC}
 & $0.0884$
 \\
\hline
\end{tabular}
\par\end{centering}
\caption{
    Mean sum of squared errors over all $d$ components of estimates of the posterior mean for the logistic regression model, formed using various algorithmic approaches as described in the main text, during an approximate wall-clock time equal to $200{,}000$ times that required to compute a single partial likelihood $f_{j}$.
	All values computed over $25$ replicates, with the lowest value printed in bold.
}
\label{tab:logRegMSE}
\end{table}

For each algorithmic setting, we ran the corresponding sampler $25$ times.
To compare the resulting estimators of the posterior mean we computed the mean squared error of each of the $d$ components of the posterior mean, summing these to obtain a `mean sum of squared errors'.

Table~\ref{tab:logRegMSE} compares the values obtained by each algorithm after an approximate wall-clock time equal to $200{,}000$ times the time taken to compute a single partial likelihood $f_{j}$.
Accounting for latency in the abstracted distributed setting described above, the GCMC approach is able to generate $5000$ approximate posterior samples during this time, spending $50\%$ of time on likelihood computations.
In contrast, a direct MCMC approach generates $9523$ samples, but would only spend $4.8\%$ of the time on likelihood computations, with the remainder lost due to latency.

The result is that the estimators generated by GCMC for appropriately-chosen $\lambda$ exhibit lower mean sums of squared errors: we conduct many more accept/reject steps in each round of inter-node communication than if we were to directly target $\pi$, and so it becomes possible to achieve faster mixing of the $Z$-chain (and a better estimator) compared to such a direct approach.
This may be seen when comparing the effective sample size (ESS) of each chain, where we estimate this via the `batch means' approach of \citet{VatsEtAl2019}: we find that the average ESS of the direct MCMC chains is only $1111$, while depending on the choice of $\lambda$, the shorter GCMC chains have average ESS values between $1327$ and $4577$.

Despite being unaffected by latency and therefore allowing many more samples to be drawn, the embarrassingly parallel approaches (CMC, NDPE, WRS) perform poorly compared to GCMC.
This is particularly true of the nonparametric density product estimation (NDPE) method of \citet{NeiswangerEtAl2014}: while asymptotically exact even in non-Gaussian settings, the resulting estimator is based on kernel density estimators and is not effective in this high-dimensional setting. 

\begin{figure}
	\centering
	\includegraphics{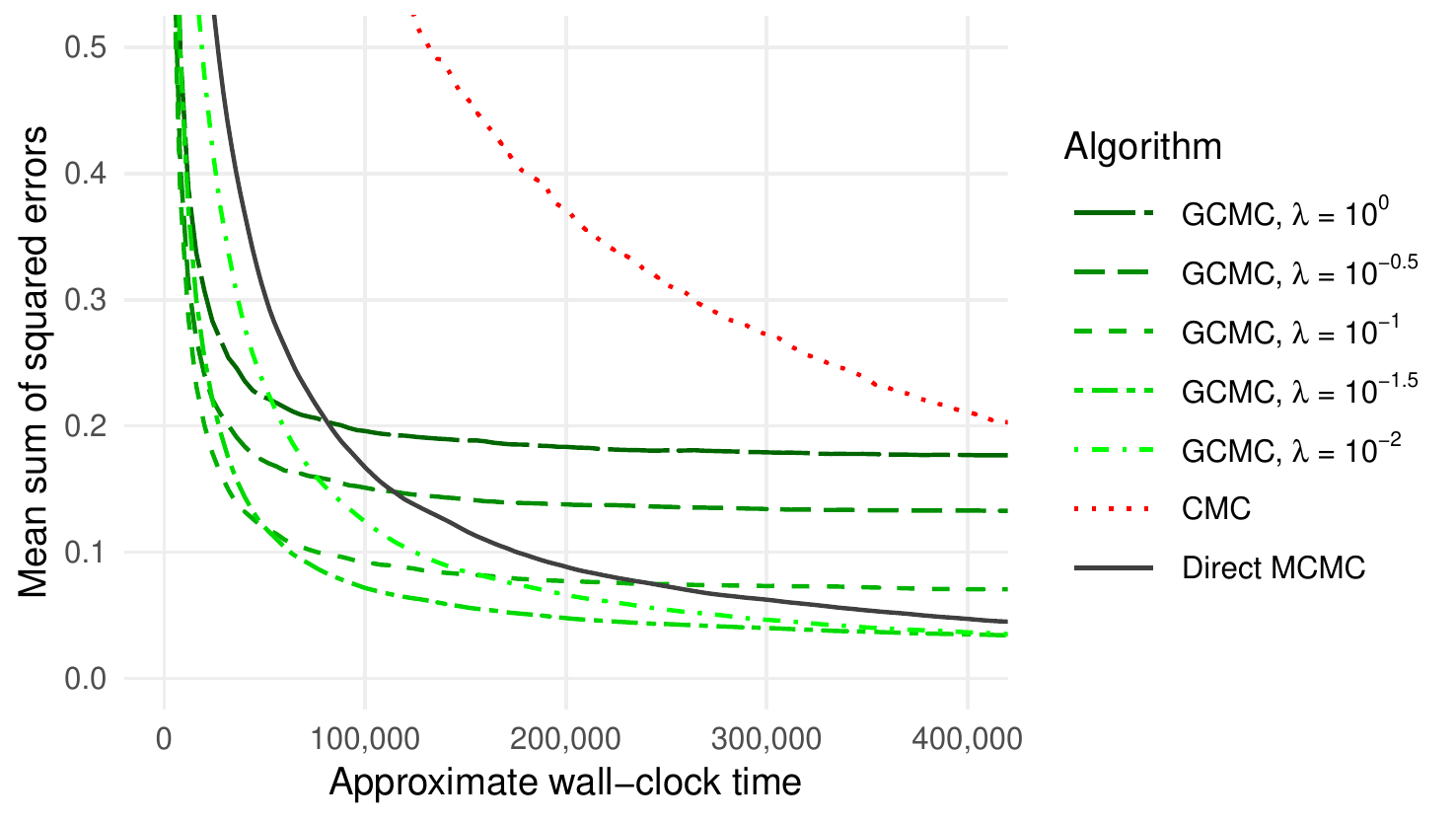}
	\caption{
    Mean sum of squared errors over all $d$ components of estimates of the posterior mean for the logistic regression model, formed using various algorithmic approaches as described in the main text.
    Values plotted against the approximate wall-clock time, relative to the time taken to compute a single partial likelihood term.
    All values computed over $25$ replicates.
    }
	\label{fig:logRegMSE}
\end{figure}

Figure~\ref{fig:logRegMSE} shows the mean sums of squared errors as a function of the approximate wall-clock time (for simplicity we include only the best-performing of the three embarrassingly parallel methods, omitting the results for NDPE and WRS).
We see that for large enough $\lambda$, the GCMC estimators $\pi_\lambda^N(\Id)$ exhibit rather lower values than the corresponding CMC and `direct' MCMC estimators.
As the number of samples used grows, the squared bias of these estimators begins to dominate, and so smaller $\lambda$ values result in lower mean squared errors.
As $\lambda$ becomes smaller the autocorrelation of the resulting $Z$-chain increases; indeed we found that for $\lambda$ too small, the GCMC estimator $\pi_\lambda^N(\Id)$ will always have a greater mean squared error than the `direct' MCMC estimator, no matter how much time is used.
Of course, since an MCMC estimator formed by directly targeting $\pi$ is consistent in $N$, given sufficient time such an estimator will always outperform estimators formed using GCMC, which are biased for any $\lambda$.
However, in many practical big data settings it may be infeasible to draw large numbers of samples using the available time budget.

\subsubsection{Sequential Monte Carlo}

We also applied the SMC procedure to this logistic regression model.
While we found that the SMC approach was most effective in lower-dimensional settings (see Section~\ref{subsec:gaussianSMCexample} of the supplement for a simple example) in which it is less computationally expensive, the SMC procedure can be more widely useful as a means of `refining' the estimator formed using a single $\lambda$ value, as discussed in Section~\ref{sec:SMCimplementation}.

We used $N = 1250$ particles, initialising the particle set by thinning the chain generated by the Metropolis-within-Gibbs procedure with $\lambda = 10^{-1}$.
To generate a sequence of subsequent $\lambda$ values we used the adaptive procedure of \citet{ZhouEtAl2016}, using tuning parameter $\text{CESS}^\star = 0.98$.
For the Markov kernels $M_{p}$ we used Metropolis-within-Gibbs kernels as previously, with each update of $X_j$ given $Z$ comprising $k = 50$ iterations of a random walk Metropolis kernel.

The estimator $\pi_{\lambda_{0}}^{N}(\Id)$ formed using the initial particle set was found to have a mean sum of squared errors of $0.0692$.
After a fixed number of iterations ($n = 100$) the resulting SMC estimate exhibited a mean sum of squared errors of $0.0418$; this represents a decrease of $40\%$, and has the benefit of avoiding the need to carefully specify a single value for $\lambda$.

Used alone, the bias correction procedure of  Section~\ref{subsec:biasCorrection} was found to perform best in lower-dimensional settings (as in Section~\ref{subsec:gaussianSMCexample} of the supplement); here, it resulted in a mean sum of squared errors of $0.0682$ after $100$ iterations.
However, improved results were obtained using the stopping rule we propose in Section~\ref{subsec:stoppingRule} of the supplement (with stopping parameter $\kappa = 15$), which is based on our proposed bias correction procedure.
The estimator selected by this stopping rule, which automatically determines when to terminate the algorithm, obtained a mean sum of squared errors of $0.0367$, a decrease of $47\%$ from the estimator generated using the initial particle set.

\section{Conclusion}
\label{sec:conclusion}

We have presented a new framework for sampling in distributed settings, demonstrating its application on some illustrative examples in comparison to embarrassingly parallel approaches.
Given that our proposed approach makes no additional assumptions on the form of the likelihood beyond its factorised form as in \eqref{eq:piOriginal}, we expect that our algorithm will be most effective in those big data settings for which approximate Gaussianity of the likelihood contributions may not hold.
These may include high-dimensional settings, for which some subsets of the data may be relatively uninformative about the parameter.
In such cases the likelihood contributions may be highly non-Gaussian, so that the consensus Monte Carlo approach of averaging across chains results in estimates of high bias; simultaneously, the high dimensionality may preclude the use of alternative combination techniques (e.g.\ the use of kernel density estimates).

This framework may be of use in serial settings. As previously noted, the contemporaneous work of \citet{VonoEtAl2019} presents an example in which the use of an analogous framework (with $b=1$) results in more efficient simulation than approaches that directly target the true posterior.
Our proposed SMC sampler implementation and the associated bias correction technique may equally be applied to such settings, reducing the need to specify a single value of the regularisation parameter $\lambda$.

There is potential for further improvements to be made to the procedures we present here.
In the SMC case for example, while our proposed use of weighted least squares as a bias correction technique is simple, non-linear procedures \citep[such those proposed in an ABC context by][]{BlumFrancois2010} may provide a more robust alternative, with some theoretical guarantees.
We also stress that the MCMC and SMC procedures presented here constitute only two possible approaches to inference within the instrumental hierarchical model that we propose, and there is considerable scope for alternative sampling algorithms to be employed within this global consensus framework.

\bigskip
\begin{center}
{\large\bf SUPPLEMENTARY MATERIAL}
\end{center}

\begin{description}

\item[Supplement to `Global Consensus Monte Carlo':] ``GCMCsupplement.pdf'' includes supplementary material covering implementation considerations, theoretical analysis for a simple model, and heuristic procedures for the proposed SMC sampler (with numerical demonstrations).

\item[R code for examples:] ``GCMCexamples.zip'' contains R scripts used to generate the numerical results and figures presented here and in the supplement.

\end{description}

\bibliographystyle{apalike}

\spacingset{1}
\bibliography{refs}
\end{document}


\def\spacingset#1{\renewcommand{\baselinestretch}%
{#1}\small\normalsize} \spacingset{1}


\if0\blind
{
  \title{\bf Supplement to `Global Consensus Monte Carlo'}
  \author{Lewis J.\ Rendell\thanks{
          The authors gratefully acknowledge the support of The Alan Turing Institute under the EPSRC grant EP/N510129/1, the Lloyd's Register Foundation--Alan Turing Institute Programme on Data-Centric Engineering, and the EPSRC under grants EP/M508184/1, EP/R034710/1 and EP/T004134/1.}\hspace{.2cm}\\
    Department of Statistics, University of Warwick\\
    and \\
    Adam M.\ Johansen \\
    Department of Statistics, University of Warwick \\
	and \\
	Anthony Lee \\
	School of Mathematics, University of Bristol \\
	and \\
	Nick Whiteley \\
	School of Mathematics, University of Bristol \\}
  \maketitle
} \fi

\if1\blind
{
  \bigskip
  \bigskip
  \bigskip
  \begin{center}
    {\LARGE\bf Global Consensus Monte Carlo}
\end{center}
  \medskip
} \fi

\vfill

\newpage
\spacingset{1.5} 

\section{Considerations when implementing the Metropolis-within-Gibbs sampler}

\subsection{Repeated MCMC kernel iterations}
\label{subsec:repeatedMCMCkernels}

To analyse the effects of communication latency on our approach, we consider an abstracted distributed setting:
\begin{itemize}
	\item Let $\ell$ represent the approximate wall-clock time required to compute each $f_j(z)$ for a given $z \in \sfE$, which we here assume is independent of $j$ for simplicity.
	\item Let the communication latency be $C$; for the purposes of this analysis we shall consider the additional time taken due to bandwidth restrictions to be negligible.
	\item Assume also that the time taken to compute the prior $\mu$, and the global consensus transition densities $K_j^{(\lambda)}$, is negligible.
\end{itemize}

In an MCMC approach directly targeting the full posterior, each accept/reject step requires communication to and from each node in order to compute the posterior density $\pi(z)$ at the proposed $z \in \sfE$.
Assuming that the likelihood contributions of each block may be computed synchronously, the time taken by each iteration of such an algorithm is therefore approximately $\ell + 2C$.

Within our proposed global consensus approach computations of the full conditional densities of each $X_j$, and of $Z$, may each occur on a single node.
Suppose that the Markov kernels $P_{j,z}^{(\lambda)}$ comprise $k$ iterations of an MCMC kernel leaving $\pi_{\lambda}(x_j \mid z)$ invariant.
Then, under the same assumptions of synchronous computation, a single iteration of our Metropolis-within-Gibbs algorithm (generating one new value of the $Z$-chain) requires a time of approximately $k \ell + 2C$.

The consequence is that, while our proposed approach would generally generate fewer samples per unit of wall-clock time, the proportion of time spent on likelihood computation (rather than communication) may be made far greater: $k\ell / (k\ell + 2C)$ for global consensus Monte Carlo, versus  $\ell / (\ell + 2C)$ for the `direct' MCMC approach.
This may be especially important when the latency $C$ is large compared to the likelihood computation time $\ell$; by choosing the number of MCMC kernel applications $k$ to be sufficiently large, the resulting sampler will spend a greater proportion of time exploring the state space, and may therefore exhibit faster mixing (with respect to wall-clock time). This approach may be particularly useful for high-dimensional settings, in which constructing a well-mixing MCMC kernel can be difficult.

The `local' application of MCMC kernels in our framework may also allow a wider range of such kernels to be computationally feasible than in a direct approach.
For example, in the case where the state space $\sfE$ is multi-dimensional, one may wish to use `componentwise' proposals, in which new values are proposed for each individual component (or collection thereof), with all others held fixed.
Updating each component in turn may be infeasible in a direct MCMC approach, due to the communication latency involved in computing the acceptance probability for each proposed value.
Similarly, our proposed framework may also be beneficial when using adaptive MCMC algorithms in distributed settings.
In a direct MCMC approach, adapting the choice of proposal distribution to the target $\pi$ may be slow (in the sense of wall-clock time) due to the communication required in each accept/reject step.
Within the global consensus framework, for which the acceptance probabilities required by the MCMC kernels $P_{j,z}^{(\lambda)}$ may be computed locally, adaptation may occur faster, which may contribute to better mixing.

Our analysis of the Gibbs sampler setting in Section~\ref{sec:Gaussians} may be informative about the more general Metropolis-within-Gibbs setting, in cases where each $P_{j,z}^{(\lambda)}$ comprises enough MCMC iterations $k$ to exhibit good mixing.

\subsection{Choosing the Markov transition densities}
\label{subsec:choosingTransitionDensities}

Depending on $\mu$ and $f_{1:b}$, appropriate choices of $K_{j}^{(\lambda)}$ may enable direct sampling from some of the full conditional distributions of $Z$ and $X_{1:b}$.
For example within \eqref{eq:zFullConditional}, each $K_{j}^{(\lambda)}(\cdot, x_{j})$ is a pseudo-likelihood for $Z \sim \mu$, and so if $\mu$ is conjugate to $K_{j}^{(\lambda)}(\cdot, x_{j})$ for each $j \in \{1,\ldots,b\}$, then the conditional distribution of $Z$ given $X_{1:b}$ will be from the same family as $\mu$.
Similarly, one might choose for each $K_{j}^{(\lambda)}(z, \cdot)$ a conjugate prior for the partial likelihood terms $f_j$, so that the conditional distribution \eqref{eq:xFullConditional} of each $X_j$ given $\mathbf{y}_j$ and $Z$ is from the same family as $K_{j}^{(\lambda)}(z, \cdot)$.

It may also be appropriate to choose the Markov transition densities to have relative scales comparable to those of the corresponding partial likelihood terms.
To motivate this consider a univariate setting in which the partial likelihood terms are Gaussian, so that we may write $f_{j}(z) = \dnorm(\mu_{j};z,\sigma_{j}^{2})$ for each $j \in \{1,\ldots,b\}$.
Suppose one uses Gaussian transition densities $K_{j}^{(\lambda)}(z,x) = \dnorm(x;z,c_{j}\lambda)$, where $c_{1}, \dots, c_{b}$ are positive values controlling the relative strengths of association between $Z$ and the local variables $X_{1:b}$.
As seen in \eqref{eq:smoothedPosterior}, in the approximating density $\pi_{\lambda}$ the partial likelihood terms $f_j$ are replaced by smoothed terms \eqref{eq:smoothedLikelihood}, in this case given by
\begin{equation}
f_{j}^{(\lambda)}(z) \propto \dnorm(\mu_{j};z,\sigma_{j}^{2} + c_{j}\lambda).
\label{eq:smoothedLikelihoodGaussian}
\end{equation}
The resulting smoothed posterior density is presented as \eqref{eq:pi_lambda} in Section~\ref{sec:Gaussians}, where this setting is further explored.

In this case, the role of $\lambda$ may be seen as `diluting' or downweighting the contribution of each partial likelihood to the posterior distribution $\pi_{\lambda}$.
A natural choice is to take $c_j \propto \sigma_{j}^{2}$, so that the dilution of each $f_j$ is in proportion to the strength of its contribution to $\pi$. In this case \eqref{eq:smoothedLikelihoodGaussian} becomes
\[
f_{j}^{(\lambda)}(z) \propto \dnorm(\mu_{j};z,(1 + c\lambda)\sigma_{j}^{2})
\]
for some constant $c$.
The relative strengths of contribution of the $f_{1:b}$ are thereby preserved in the posterior density $\pi_{\lambda}$.

A particular case of interest in that in which the blocks of data $\mathbf{y}_j$ differ in size.
If each observation $y_{\ell}$ has a likelihood contribution of the form $\dnorm(y_{\ell};z,\sigma^{2})$, then the $j$th partial likelihood may be expressed as $f_{j}(z) \propto \dnorm(\bar{y}_{j};z,\sigma^{2}/n_{j})$, where $n_j$ is the number of data in the $j$th block and $\bar{y}_{j}$ is their mean.
Taking $c_j \propto 1/n_{j}$, the smoothed partial likelihood \eqref{eq:smoothedLikelihoodGaussian} becomes
\[
f_{j}^{(\lambda)}(z) \propto \dnorm(\bar{y}_{j};z,(\sigma^{2} + c\lambda)/n_{j})
\]
for some $c$, so that the information from each observation is diluted in a consistent way.
Motivated by Bayesian asymptotic arguments, we suggest that this scaling of the regularisation parameter in inverse proportion to the relative block sizes may be beneficial in more general settings.

The effect of such choices on the MCMC algorithm is most readily seen by considering the improper uniform prior $\mu(z)\propto1$ (a Gaussian prior is considered in Section~\ref{sec:Gaussians}).
Taking $K_{j}^{(\lambda)}(z,x)=\dnorm(x;z,c_{j}\lambda)$, the conditional density of $Z$ given $X_{1:b}$ is
\[
\tilde{\pi}_{\lambda}(z\mid x_{1:b}) =
\dnorm \left( z; \frac{\sum_{j=1}^{b} x_{j}/c_{j}}{\sum_{j=1}^{b} 1/c_{j}},
\frac{\lambda}{\sum_{j=1}^{b} 1/c_{j}} \right).
\]
Therefore, when updating $Z$ given the local variables' current values $x_{1:b}$, the choice of $c_{1:b}$ dictates the relative influence of each such value.
For example, we might expect the local variables corresponding to larger blocks to be more informative about the distribution of $Z$, which further justifies choosing $c_{1:b}$ to be inversely proportional to the block sizes.

In a multidimensional setting, one could control the covariance structure of each $X_{j}$ given $Z$ by using transition densities of the form $\dnorm(x;z,\lambda\Psi_{j})$, where $\Psi_{1:b}$ are positive semi-definite matrices.
By a similar Gaussian analysis, one could preserve the relative strengths of contribution of the partial likelihood terms by choosing for each $\Psi_{j}$ an approximation of the covariance matrix of $f_j$.

\section{Theoretical analysis for a simple model}
\label{sec:Gaussians}

\subsection{Inferring the mean of a normal distribution}
\label{sub:Inferring-the-mean}

To study the theoretical properties of our algorithm, we here consider a simple model where the goal is to infer the mean of a normal distribution.
While our approach does not require the distribution $\pi$ to be approximately Gaussian, the behaviour of our algorithm in this simple setting is particularly amenable to analysis.
The results here may also be indicative of performance for regular models with abundant data due to the Bernstein--von Mises theorem \citep[see, e.g.,][Chapter 10]{Vaart2000}.

Let $\mu(z) = \dnorm(z;\mu_{0},\sigma_{0}^{2})$, and for each $j\in\{1,\dots,b\}$ let $f_{j}(z) = \dnorm(\mu_{j};z,\sigma_{j}^{2})$ and $K_{j}^{(\lambda)}(z,x) = \dnorm(x;z,c_{j}\lambda)$, following Section~\ref{subsec:choosingTransitionDensities}.
We obtain
\begin{equation}
\pi_{\lambda}(z) = \dnorm \left(z; \delta^2_\lambda \left[ \frac{\mu_{0}}{\sigma_{0}^{2}} + \sum_{j=1}^{b} \frac{\mu_{j}}{\sigma_{j}^{2} + c_{j}\lambda} \right], \delta^2_\lambda \right),
\quad
\delta^2_\lambda = \left ( \frac{1}{\sigma_{0}^{2}} + \sum_{j=1}^{b} \frac{1}{\sigma_{j}^{2} + c_{j}\lambda} \right )^{-1}
\label{eq:pi_lambda},
\end{equation}
and $\pi(z)$ can be recovered by taking $\lambda=0$ in \eqref{eq:pi_lambda}.
The corresponding full conditional densities for \eqref{eq:pi_lambda} are
\[
\tilde{\pi}_{\lambda}(x_{j}\mid z)
=
\dnorm\left(x_{j};
\frac{\sigma_{j}^{2}z+c_{j}\lambda\mu_{j}}{\sigma_{j}^{2}+c_{j}\lambda},
\frac{c_{j}\lambda\sigma_{j}^{2}}{\sigma_{j}^{2}+c_{j}\lambda}\right)
\]
and
\[
\tilde{\pi}_{\lambda}(z\mid x_{1:b})
=
\dnorm\left(z;
\tilde{\delta}^{2}_{\lambda}
\left[\frac{\mu_{0}}{\sigma_{0}^{2}} + \sum_{j=1}^{b}\frac{x_{j}}{c_{j}\lambda} \right],
\tilde{\delta}^{2}_{\lambda}
\right),
\quad
\tilde{\delta}^{2}_{\lambda} = \left(\frac{1}{\sigma_{0}^{2}} + \sum_{j=1}^{b}\frac{1}{c_{j}\lambda} \right)^{-1},
\]
and we consider the case where $M_{1}^{(\lambda)}$ and $M_{2}^{(\lambda)}$ as defined in \eqref{eq:M1kernel}--\eqref{eq:M2kernel} are Gibbs kernels.
That is, we consider the case in which we draw samples exactly from the full conditional distributions \eqref{eq:xFullConditional}--\eqref{eq:zFullConditional}.
We choose this setting to facilitate analysis of the resulting Markov chain, but these results may be informative about more general Metropolis-within-Gibbs settings in which well-mixing Markov kernels are used (e.g. those comprising multiple MCMC kernel iterations, as described in  Section~\ref{subsec:repeatedMCMCkernels}).

Since $X_{1:b}$ are conditionally independent given $Z$ and can therefore be updated simultaneously given $Z$, Algorithm~\ref{alg:gcmcMCMC} may be viewed analytically as a Gibbs sampler on two variables: $Z$ and $X_{1:b}$.
For any two-variable Gibbs Markov chain, each of the two `marginal' chains (the sequences of states for each of the two variables) is also a Markov chain. In this setting we may therefore consider the $Z$-chain with transition kernel
\begin{equation}
M_{12}^{(\lambda)}(z,\diff z') = 
\left[ \int_{\sfE^{b}} \tilde{\pi}_{\lambda}(z'\mid x_{1:b}) \prod_{j=1}^{b} \tilde{\pi}_{\lambda}(x_{j} \mid z) \diff x_{1:b} \right] \diff z' .
\label{eq:gibbsZchainKernel}
\end{equation}
Observing that $\tilde{\pi}_{\lambda}(z' \mid x_{1:b})$ depends on $x_{1:b}$ only through the sum $\sum_{j=1}^{b}x_{j}/c_{j}$, one can thereby show that the $Z$-chain defined by \eqref{eq:gibbsZchainKernel} is an AR(1) process.
Specifically, we have
\[
Z^{i} = C + \alpha Z^{i-1} + \epsilon_{i}
,\qquad k>0,
\]
where
\[
\alpha \coloneqq
\tilde{\delta}^{2}_{\lambda} \sum_{j=1}^{b} \frac{\sigma_{j}^{2}}{c_{j}\lambda(\sigma_{j}^{2}+c_{j}\lambda)}
, \qquad
C \coloneqq
\tilde{\delta}^{2}_{\lambda} \left( \frac{\mu_{0}}{\sigma_{0}^{2}} + \sum_{j=1}^{b} \frac{\mu_{j}}{\sigma_{j}^{2} + c_{j}\lambda} \right),
\]
and the $\epsilon_{i}$ are i.i.d.\ zero-mean normal random variables, with variance
\[
\tilde{\delta}^{2}_{\lambda} \left( 1 + \tilde{\delta}^{2}_{\lambda}
\sum_{j=1}^{b} \frac{\sigma_{j}^{2}}{c_{j}\lambda(\sigma_{j}^{2}+c_{j}\lambda)} \right).
\]
It follows that the autocorrelation of lag $k$ is given by $\alpha^{k}$ for $k\geq0$, and that $\alpha \to 1$ as $\lambda \to 0$.

\subsection{Asymptotic bias and variance with $n$ observations}
\label{subsec:gaussianAsymptotics}

We now consider the setting of Section~\ref{sub:Inferring-the-mean}, making the number of observations $n$ explicit.
In particular, for some $z^* \in \reals$ consider realisations $y_{1:n}$ of i.i.d.\  $\dnorm(z^*,\sigma^2)$ random variables, grouped into $b$ blocks.
For simplicity, assume that $b$ divides $n$, that each block contains $n/b$ observations, and that the observations are allocated to the blocks sequentially, so that the $j$th block comprises those $y_{\ell}$ where $\ell \in B_{j} \coloneqq \{(j-1)n/b+1,\dots,jn/b\}$.
Then
\[
f_{j}(z) =
\prod_{\ell\in B_{j}} \dnorm(y_{\ell};z,\sigma^{2})
\propto \dnorm \left( \frac{b}{n} \sum_{\ell\in B_{j}} y_{\ell}; z, \frac{b}{n}\sigma^{2} \right).
\]
Since the blocks are of equal size in this case, so that each partial likelihood is of the same scale, we consider using $K_{j}^{(\lambda)}(z,x) = \dnorm(x;z,\lambda)$ for each $j$. From \eqref{eq:pi_lambda}, we obtain
\begin{equation}
\pi_{\lambda}(z)
= \dnorm \left(z; \delta_\lambda^2 \left[ \frac{\mu_{0}}{\sigma_{0}^{2}} + \frac{n\bar{y}}{\sigma^{2}+n\lambda/b} \right] , \delta_\lambda^2 \right)
, \quad \delta_\lambda^2 = \left ( \frac{1}{\sigma_{0}^{2}} + \frac{n}{\sigma^{2}+n\lambda/b} \right )^{-1}.
\label{eq:gaussianPiLambda}
\end{equation}

Denoting the identity function on $\reals$ by $\Id$, we consider an estimator $\pi_\lambda^N(\Id)$ of the posterior first moment $\pi(\Id)$, as formed according to \eqref{eq:gcmcEstimator}.
We analyse its mean squared error using the bias--variance decomposition \eqref{eq:gcmc_mse}.
The bias is
\begin{equation}
\pi_\lambda(\Id) - \pi(\Id) =
\frac{n^{2}\left( \lambda/b \right) \sigma_{0}^{2} \left( \mu_{0}-\bar{y} \right)}{\left( \sigma^{2} + n\sigma_{0}^{2} \right) \left( \sigma^{2} + n\sigma_{0}^{2} + n\lambda/b \right)}.
\label{eq:gaussianBias}
\end{equation}
To assess the variance of $\pi_\lambda^N(\Id)$, we consider the associated \emph{asymptotic} variance,
\begin{equation}
\lim_{N \to \infty} N \var(\pi^N_\lambda(\varphi))
= \var(\varphi(Z_{0})) \left[ 1 + 2\sum_{k=1}^{\infty} \corr (\varphi(Z_{0}), \varphi(Z_{k})) \right]
,\qquad Z_{0} \sim \pi_{\lambda},
\label{eq:asymptoticVariance}
\end{equation}
for $\varphi$ square-integrable w.r.t.\ $\pi_\lambda$.
As discussed earlier the $Z$-chain is an AR(1) process, and the autocorrelations are entirely determined by the autoregressive parameter
\[
\alpha = \frac{n\sigma^{2}\sigma_{0}^{2}}{\left( \sigma^{2}+n\lambda/b \right) \left(n\sigma_{0}^{2}+n\lambda/b \right)},
\]
from which one can find that the asymptotic variance for $\varphi = \Id$ is
\begin{equation}
\frac{\sigma_{0}^{2} \left( \sigma^{2}+n\lambda/b \right) \left[ \left( n\lambda/b \right)^{2} + \left( \sigma^{2}+n\sigma_{0}^{2} \right) \left( n\lambda/b\right ) + 2n \sigma^{2}\sigma_{0}^{2} \right]}{\left( n\lambda/b \right) \left(\sigma^{2}+n\sigma_{0}^{2}+n\lambda/b \right)^{2}}.
\label{eq:gaussianAsymptoticVariance}
\end{equation}
Following the definition \eqref{eq:asymptoticVariance} of this asymptotic variance, dividing this expression by $N$ gives an approximation of the variance term in \eqref{eq:gcmc_mse} for large $N$.

As a caveat to this and the following analysis, estimation of the mean in Gaussian settings may not accurately reflect what happens in more complex settings.
For example, if one uses an improper uniform prior then $\pi_{\lambda}(\Id)$ is equal to $\pi(\Id)$ for all $\lambda$, as seen in \eqref{eq:gaussianBias} with $\sigma_{0}^{2}\to\infty$; this will not be true in general.

One may also note that in this Gaussian setting, the variance of $\pi_{\lambda}$ will always exceed the variance of the true target $\pi$, since the variance expression in \eqref{eq:gaussianPiLambda} is an increasing function of $\lambda$.
The effect is that estimation of the posterior variance in Gaussian settings is likely to result in positive bias, and confidence intervals for $\pi(\Id)$ may be conservative.
This, of course, simply reflects the fact that marginally the instrumental model can be viewed as replacing the original likelihood with a smoothed version as shown in \eqref{eq:smoothedPosterior}.

\subsection{Asymptotic optimisation of $\lambda$ for large $N$}
For fixed $n$, we consider the problem of choosing $\lambda$ as a function of chain length $N$ so as to minimise the mean squared error of the posterior mean estimator.
This involves considering the contributions of the bias and variance to the mean squared error \eqref{eq:gcmc_mse} in light of \eqref{eq:gaussianBias} and \eqref{eq:gaussianAsymptoticVariance}.
Intuitively, with larger values of $N$, smaller values of $\lambda$ can be used to reduce the bias while keeping the variance small. Defining $B(\lambda)$ to be the bias as given in \eqref{eq:gaussianBias}, we see that as $\lambda \to 0$,
\[
\frac{B(\lambda)}{\lambda} \to \frac{n^{2}\sigma_{0}^{2} \left( \mu_{0}-\bar{y} \right)}{b\left(\sigma^{2}+n\sigma_{0}^{2} \right)^{2}} \eqqcolon B_{\star}.
\]
Similarly, denoting by $V(\lambda)$ the asymptotic variance \eqref{eq:gaussianAsymptoticVariance}, we see that
\[
\lambda V(\lambda) \to \frac{2b\sigma^{4}\sigma_{0}^{4}}{\left( \sigma^{2}+n\sigma_{0}^{2} \right)^{2}} \eqqcolon V_{\star}.
\]
For small $\lambda$, the mean squared error of the estimate is given approximately by
\[
\E \! \left[ \left( \pi_\lambda^N(\Id) - \pi(\Id) \right)^2 \right] \approx \left( \lambda B_{\star} \right)^{2} + \frac{1}{N}\frac{V_{\star}}{\lambda},
\]
which may be shown to be minimised when
\[ \lambda^{3}
= \frac{V_{\star}}{2B_{\star}^{2}N}
= \frac{b^{3}\sigma^{4} \left( \sigma^{2}+n\sigma_{0}^{2} \right)^{2}}{n^{4}N \left( \mu_{0}-\bar{y} \right)^{2}}.
\]
We see that, for a fixed number of data $n$, we should scale $\lambda$ with the number of samples $N$ as $\mathcal{O}(N^{-1/3})$. The corresponding minimal mean squared error, in which the contribution of the variance is twice that of the squared bias, may be shown to behave as $\mathcal{O}(N^{-2/3})$.

Note that in this example, all dependence on $\lambda$ and $b$ in the smoothed likelihood \eqref{eq:gaussianPiLambda} is through their ratio $\lambda/b$.
The result is that splitting the data into more blocks has the same effect as reducing $\lambda$, and so these results may be adapted to consider optimisation of the ratio $\lambda/b$.
This relationship may not be representative of these variables' behaviour in other models; but in cases where Bernstein--von Mises arguments hold, such results may be useful in settings where the number of blocks $b$ may be chosen by the practitioner.

\subsection{Posterior consistency and coverage of credible intervals as $n \to \infty$}

We now consider the behaviour of the algorithm as the number of data $n$ tends to infinity.
Recalling that we assume the true parameter value to be $z^* \in \reals$, we may consider the consistency of the posterior distribution \eqref{eq:gaussianPiLambda} by treating the data $Y_{1:n}$ as random.
We denote their mean by $\bar{Y}_{n}$, which is normally distributed with mean $z^*$ and variance $\sigma^{2}/n$.
We shall also consider allowing $\lambda$ and $b$ to vary with $n$; making this explicit in the notation, \eqref{eq:gaussianPiLambda} becomes
\[
\pi_{\lambda_{n}}(z)
= \dnorm \left( z;\mu_{(n)},\delta_{(n)}^{2} \right)
= \dnorm \left( z;\delta_{(n)}^{2}\xi_{(n)},\delta_{(n)}^{2} \right),
\]
where $\mu_{(n)} = \delta_{(n)}^{2}\xi_{(n)}$, and
\[
\xi_{(n)} = \frac{\mu_{0}}{\sigma_{0}^{2}} + \frac{n\bar{Y}_{n}}{\sigma^{2}+n\lambda_{n}/b_{n}}
,\qquad
\delta_{(n)}^{2} = \left( \frac{1}{\sigma_{0}^{2}} + \frac{n}{\sigma^{2}+n\lambda_{n}/b_{n}} \right)^{-1}.
\]
We consider $\lambda_{n}/b_{n}= c n^{-\gamma}$ and using the fact that $\bar{Y}_{n} \almostsurelyto z^{*}$, we obtain the following convergence results for different values of $\gamma$:
\begin{center}
\begin{tabular}{cccc}
\hline
$\gamma<0$
& $\delta_{(n)}^{2}\to\sigma_{0}^{2}$
& $\xi_{(n)} \almostsurelyto \frac{\mu_{0}}{\sigma_{0}^{2}}$
& $\mu_{(n)} \almostsurelyto \mu_{0}$
\tabularnewline
\hline
$\gamma=0$
&  $\delta_{(n)}^{2}\to\left(\frac{1}{\sigma_{0}^{2}}+\frac{1}{c}\right)^{-1}$
& $\xi_{(n)} \almostsurelyto \frac{\mu_{0}}{\sigma_{0}^{2}}+\frac{z^{*}}{c}$
&
$\mu_{(n)} \almostsurelyto \left(\frac{1}{\sigma_{0}^{2}}+\frac{1}{c}\right)^{-1}\left(\frac{\mu_{0}}{\sigma_{0}^{2}}+\frac{z^{*}}{c}\right)$
\tabularnewline
\hline
$\gamma\in(0,1)$
& $n^{\gamma}\delta_{(n)}^{2}\to c$
& $n^{-\gamma}\xi_{(n)} \almostsurelyto \frac{z^{*}}{c}$
& $\mu_{(n)} \almostsurelyto z^{*}$
\tabularnewline
\hline
$\gamma=1$
& $n\delta_{(n)}^{2}\to\sigma^{2}+c$
& $n^{-1}\xi_{(n)} \almostsurelyto \frac{z^{*}}{\sigma^{2}+c}$
& $\mu_{(n)} \almostsurelyto z^{*}$
\tabularnewline
\hline
$\gamma>1$
& $n\delta_{(n)}^{2}\to\sigma^{2}$
& $n^{-1}\xi_{(n)} \almostsurelyto \frac{z^{*}}{\sigma^{2}}$
& $\mu_{(n)} \almostsurelyto z^{*}$
\tabularnewline
\hline
\end{tabular}
\end{center}
Hence, it can be seen that the posterior is consistent \citep[see, e.g.,][Chapter 1]{GhoshRamamoorthi2003} if $\gamma>0$.
Moreover, if $\gamma>1$ then $1-\alpha$ credible intervals will have asymptotically a coverage probability of exactly $1-\alpha$ due to the convergence $n \delta^2_{(n)} \to \sigma^2$.

If $\gamma \in (0,1)$ then the rate of approximate posterior contraction is too conservative, while if $\gamma = 1$ the corresponding credible intervals will be too wide by a constant factor depending on $c$.
From a practical perspective, one can consider the case where $n/b$ corresponds to the maximum number of data points that can be processed on an individual computing node.
In such a setting, letting $b_n \propto n$ is reasonable and we require in addition that $\lambda_n$ is decreasing to obtain credible intervals with asymptotically exact coverage.

\section{Heuristic procedures for the sequential Monte Carlo implementation}

\subsection{Determining a subset of estimates to use for linear regression}
\label{subsec:inclusionProcedure}

If the local linear regression approach for bias correction is used, then the practitioner must determine a value of $\lambda$ below which the dependence of $\pi_\lambda(\varphi)$ on $\lambda$ is approximately linear.
For this purpose, we propose a heuristic based on the coefficient of determination, commonly denoted $R^2$; here, this may be thought of as the proportion of the variance of the observed values of $\pi_\lambda^N(\varphi)$ that is explained by an assumed linear dependence on $\lambda$.
To define this explicitly, consider the weighted least squares fit for which \eqref{eq:biasCorrectedEstimate} is the resulting bias-corrected estimate. Extending the notation used therein, let  $\hat{\eta}^{S}_{p}$ denote the predicted value of $\eta_{p}$ under the model.
Then the coefficient of determination $R^2_{S}$ for this weighted least squares model fit may be computed as the ratio of the weighted sum of squared errors and the weighted total sum of squares.
That is,
\begin{equation}
R^2_{S}
\coloneqq 
\frac{\sum_{p \in S}(\hat{\eta}_{p}^{S} - \tilde{\eta}_{S})^2 / v_{p}}{\sum_{p \in S} (\eta_{p} - \tilde{\eta}_{S})^2 / v_{p}}
=
1 - \frac{\sum_{p \in S}(\eta_{p} - \hat{\eta}_{p}^{S})^2 / v_{p}}{\sum_{p \in S} (\eta_{p} - \tilde{\eta}_{S})^2 / v_{p}}.
\label{eq:rSquared}
\end{equation}

\begin{algorithm}
	\caption{Linear regression inclusion procedure for SMC bias-correction}
	\label{alg:gcmcSMCinclusionRule}
	Fix a decreasing sequence $(\lambda_{0}, \lambda_{1}, \dots, \lambda_{n})$ and a test function $\varphi$. \\ Set number of particles $N$.
	
	\begin{enumerate}
		\item Complete Algorithm~\ref{alg:gcmcSMC}, generating and storing estimates $\eta_p \coloneqq \pi_{\lambda_{p}}^N(\varphi)$ using the particle approximations \eqref{eq:SMCapproximation_marginal}, and estimates $v_p$ of their variances, for $p \in \{0, \dots, n\}$.
		\item Initialise set of indices of estimates to be used in regression as $S \gets \{0, \dots, n\}$.
		\item Regress $\eta_{p}$ against $\lambda_{p}$ using weighted least squares, with weights $1/v_{p}$, for $p \in S$. Compute the coefficient of determination $R^2_{S}$ according to \eqref{eq:rSquared}.
		\begin{enumerate}
			\item \label{step:recomputeRsquared} If $|S| \leq 3$, proceed to Step~\ref{step:computeBCestimate}. Otherwise, set $S' \gets S \setminus \{\min(S)\}$, and regress $\eta_{p}$ against $\lambda_{p}$ using weighted least squares, with weights $1/v_{p}$, for $p \in S'$. Compute $R^2_{S'}$ according to \eqref{eq:rSquared}.
			\item If $R^2_{S'} > R^2_{S}$, set $S \gets S'$, and return to Step~\ref{step:recomputeRsquared}. Otherwise, proceed to Step~\ref{step:computeBCestimate}.
		\end{enumerate}
		\item \label{step:computeBCestimate} Return the bias-corrected estimate $\pi^{\mathrm{BC}}_{S}(\varphi)$, computed according to \eqref{eq:biasCorrectedEstimate}.
	\end{enumerate}
	
\end{algorithm}

The heuristic procedure for determining such a subset of the estimates is presented in Algorithm~\ref{alg:gcmcSMCinclusionRule}.
After completion of the SMC sampler described in Algorithm~\ref{alg:gcmcSMC}, one conducts weighted least squares in the manner described in Section~\ref{subsec:biasCorrection}, including all values of $\lambda_{p}$ and the corresponding SMC estimates $\pi_{\lambda_{p}}^N(\varphi)$, and computing the $R^2$ value for the resulting fit.
One then re-conducts the regression, without the observation in the subset corresponding to the largest $\lambda$ value.
If this results in a greater $R^2$ value, this observation should henceforth be excluded from the least squares regression.
One continues to apply this procedure, each time repeating the regression without the observation corresponding to the highest remaining value of $\lambda$, until doing so no longer results in a model with a greater $R^2$ value than the current fit.
The regression fit at this point may then be used to compute the bias-corrected estimate for $\pi(\varphi)$, so that in \eqref{eq:biasCorrectedEstimate}, $S$ corresponds to the set of indices of the remaining $\lambda$ values.

The motivation for this approach is that if this largest $\lambda$ value is not sufficiently close to zero for $\pi_\lambda(\varphi)$ to be approximately linear in $\lambda$, then retaining the corresponding SMC estimate in the regression may result in a large proportion of the variance in the data being unexplained by a linear dependence.
By excluding the corresponding SMC estimate, one would expect the linear fit applied to the remaining estimates to better describe their variance, and therefore to have a greater $R^2$ value.

This heuristic approach has a natural online implementation, allowing a bias-corrected estimate to be computed after each step of the algorithm; we use this form within our proposed stopping rule in Section~\ref{subsec:stoppingRule}, for which it forms Step~\ref{step:inclusionRuleOnline} in each iteration of Algorithm \ref{alg:gcmcSMCstoppingRule}.
Specifically, we maintain a set of the SMC estimates to be used in the regression (and the corresponding values of $\lambda$), initialising this to be empty.
After the $p$th step of the SMC sampler, the newly-generated SMC estimate $\pi_{\lambda_{p}}^N(\varphi)$ is added to this set (with the corresponding $\lambda_{p}$).
One conducts weighted least squares on this set of estimates and then, as long as the set contains more than $3$ estimates, one proceeds in the manner described above, re-conducting the regression without the observation in the set corresponding to the highest value of $\lambda$.
If this results in a fit with a higher $R^2$ value, then the omitted SMC estimate is henceforth excluded from the set used for regression, and this step is repeated.
If not, then one terminates this procedure and proceeds to the next step of the SMC sampler.

\subsection{Stopping rule}
\label{subsec:stoppingRule}

As $\lambda$ approaches zero we expect the bias resulting from estimating $\pi(\varphi)$ by $\pi_{\lambda}^N(\varphi)$ to decrease, while the variance of the resulting estimators may increase due to poorer mixing of the associated Markov kernels.
Based on the bias--variance decomposition \eqref{eq:gcmc_mse} of the mean squared error, we here propose a procedure for determining when to terminate the SMC sampler, in order to achieve such a bias--variance trade-off.

At each stage, having computed an updated bias-corrected estimate via the online procedure described in Section~\ref{subsec:inclusionProcedure}, one may subtract this value from each of the SMC estimates generated so far in order to produce an estimate of the bias in each case.
As discussed in Section~\ref{subsec:biasCorrection}, we also have an estimate of the variance of each SMC estimate, as used in the weighted linear regression procedure.
As such, at each stage we are able to estimate the mean squared error of each SMC estimate so far generated, by squaring each estimate of the bias and adding the appropriate estimate of the variance.

The formation of these mean squared error estimates is based on, but does not exactly correspond to, the bias--variance decomposition \eqref{eq:gcmc_mse}.
For example, the particle-based SMC estimates $\pi_{\lambda}^N(\varphi)$ are not unbiased as estimators of $\pi(\varphi)$, although they are consistent in the number of particles $N$ \citep[Section~3]{BeskosEtAl2016}.
Furthermore, the bias-corrected estimate itself is not unbiased, since it is formed based on approximate local linearity rather than a true linear dependency of $\pi_{\lambda}^N(\varphi)$ on $\lambda$.
Nonetheless, the use of this heuristic approach in our proposed stopping rule has been found to work well in practice, resulting in estimates of low mean squared error; we discuss one such example in Section~\ref{subsec:gaussianSMCexample}.

Note that, since the bias-corrected estimate of $\pi(\varphi)$ is updated after each step (to take into account the most recent estimate), the estimated mean squared errors of all previous estimates may also all be updated after each step.
After each SMC estimate is generated we may therefore determine which SMC estimate, of all those generated so far, has the lowest estimated mean squared error.
We propose that, for some $\kappa$, the SMC sampler should be terminated after the same previous estimate is found to have the lowest mean squared error of all those generated so far, for $\kappa$ consecutive iterations.
Following the termination of the algorithm via the stopping rule this SMC estimate, which has been repeatedly found to have the lowest estimated MSE, may be returned as the final estimate for $\pi(\varphi)$. 

\begin{algorithm}
	\caption{Global consensus Monte Carlo: SMC algorithm with stopping rule}
	\label{alg:gcmcSMCstoppingRule}
	Fix a decreasing sequence $(\lambda_{0}, \lambda_{1}, \dots, \lambda_{n})$ and a test function $\varphi$. \\
	Set number of particles $N$ and stopping rule parameter $\kappa$. \\
	Initialise set of indices of estimates to be used in regression as $S \gets \varnothing$.

	\medskip
	
	For $p = 0, 1, \dots$ until termination:
	
	\begin{enumerate}
		\item Complete the $p$th step of Algorithm~\ref{alg:gcmcSMC}, generating and storing an estimate $\eta_p \coloneqq \pi_{\lambda_{p}}^N(\varphi)$ using the particle approximation \eqref{eq:SMCapproximation_marginal}, and an estimate $v_p$ of its variance.	
		\item \label{step:inclusionRuleOnline} Set $S \gets S \cup \{p\}.$ If $|S| > 1$:
		\begin{enumerate}
			\item Regress $\eta_{q}$ against $\lambda_{q}$ using weighted least squares, with weights $1/v_{q}$, for $q \in S$. Compute the coefficient of determination $R^2_{S}$ according to \eqref{eq:rSquared}.
			\item \label{step:recomputeRsquaredOnline} If $|S| \leq 3$, proceed to Step~\ref{step:computeBCestimateOnline}. Otherwise, set $S' \gets S \setminus \{\min(S)\}$, and regress $\eta_{q}$ against $\lambda_{q}$ using weighted least squares, with weights $1/v_{q}$, for $q \in S'$. Compute $R^2_{S'}$ according to \eqref{eq:rSquared}.
			\item If $R^2_{S'} > R^2_{S}$, set $S \gets S'$, and return to Step~\ref{step:recomputeRsquaredOnline}. Otherwise, proceed to Step~\ref{step:computeBCestimateOnline}.
		\end{enumerate}
		\item \label{step:computeBCestimateOnline} Set $m_p \gets \pi^{\mathrm{BC}}_{S}(\varphi)$, a bias-corrected estimate computed according to \eqref{eq:biasCorrectedEstimate}.
		\item Set \[
		i_p \gets \argmin_{0 \leq q \leq p} \left[ (\eta_q - m_p)^2 + v_q \right],
		\]
		which corresponds to taking the index of the SMC estimate with the lowest estimated mean squared error.
		\item If $p \geq \kappa - 1$, and $(i_{p - \kappa + 1}, \dots, i_p)$ are all equal, terminate the algorithm, returning the estimate $\eta_{i_p}$ of lowest estimated MSE (and/or $m_p$, the final bias-corrected estimate).
	\end{enumerate}
	
\end{algorithm}

This approach is described in Algorithm~\ref{alg:gcmcSMCstoppingRule}. In our simulation studies, we found that taking $\kappa = 15$ worked well in balancing robustness with the computational complexity of the resulting algorithm.

As an alternative, one may choose to return the final bias-corrected estimate, for which this approach also provides a justifiable stopping rule: repeatedly finding that the same previous estimate has the lowest MSE suggests stability in our estimates of the MSEs of each previous $\pi_{\lambda}^N(\varphi)$, and therefore in the bias-corrected estimate.
Furthermore, since we expect the biases of the estimates $\pi_{\lambda}^N(\varphi)$ to decrease as $\lambda$ approaches zero, repeatedly finding that a previous SMC estimate has the lowest MSE suggests that more recent estimates are of higher variances.
Again, this is also indicative of a stabilisation of the bias-corrected estimate, since new observations are included in the regression-based bias correction procedure with weights inversely proportional to these variances.

\section{Additional examples}
\label{sec:additionalExamples}

\subsection{SMC bias correction: Gaussian example}
\label{subsec:gaussianSMCexample}

\begin{figure}
	\centering
	\begin{subfigure}[t]{.48\linewidth}
		\centering
		\includegraphics[width=.9\linewidth]{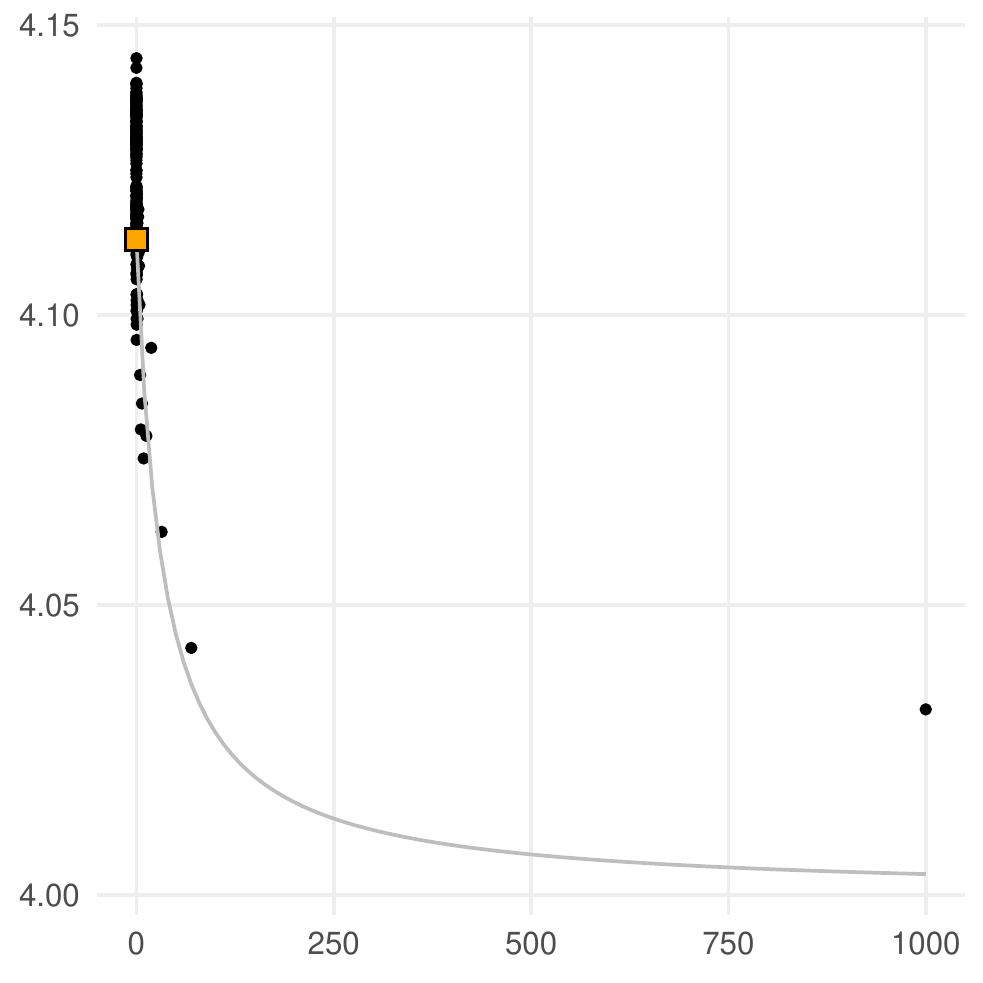}
		\caption{All estimates $\pi_{\lambda}^{N}(\Id)$, and the true value of $\pi_{\lambda}(\Id)$ as a function of $\lambda$ (solid grey line).}
		\label{fig:smcNormal_all}
	\end{subfigure}
	\quad
	\begin{subfigure}[t]{.48\linewidth}
		\centering
		\includegraphics[width=.9\linewidth]{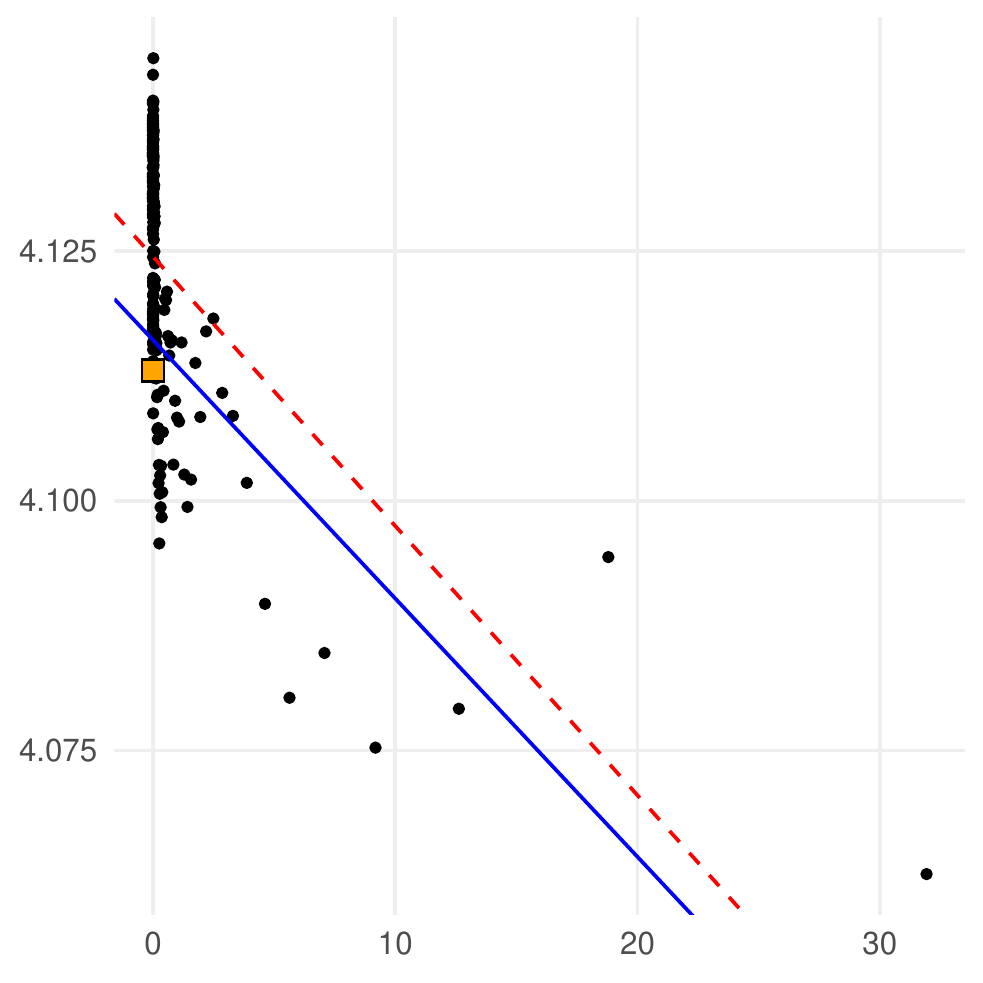}
		\caption{Subset of estimates used for local linear regression, and linear fits using least squares: weighted (solid blue line) and unweighted (dashed red line).}
		\label{fig:smcNormal_subset}
	\end{subfigure}
	\caption{
		Estimates $\pi_{\lambda}^{N}(\Id)$ plotted against $\lambda$, as obtained at each step of a single run of the SMC sampler for the Gaussian toy example. 
		The orange square indicates the true value of $\pi(\Id) \approx 4.113$.
	}
	\label{fig:smcNormal}
\end{figure}

To demonstrate the SMC bias correction technique described in Section~\ref{subsec:biasCorrection}, we consider a univariate Gaussian model of the form described in Section~\ref{sec:Gaussians}, with the aim of estimating the posterior first moment $\pi(\Id)$.
We consider a case with $b = 32$, taking $f_{j}(z)=\dnorm(\mu_{j};z,1)$ for $j \in \{1,\dots,b\}$, with the values $\mu_{j}$ drawn independently from a normal distribution with mean $4$ and variance $1$.
For the Markov transition kernels we use $K_{j}^{(\lambda)}(z,x)=\dnorm(x;z,\lambda)$.
For the purposes of illustrating the local linear regression approach to bias correction we consider the (quite concentrated) prior density $\mu(z)=\dnorm(z;4,1)$.
In this case, we see that the dependence of $\pi_{\lambda}(\Id)$ on $\lambda$ is highly non-linear on the range $0 < \lambda < 1000$ (see Figure~\ref{fig:smcNormal_all}).

We constructed an SMC sampler using $N = 2500$ particles; we used sequences of $\lambda$ values beginning with $\lambda_{0} = 1000$, with subsequent values determined adaptively according to the procedure proposed by \citet{ZhouEtAl2016}, for which we used parameter $\text{CESS}^\star = 0.95$.
For the purposes of illustrating the bias correction technique we here consider sequences of $\lambda$ values of fixed length $n = 200$; we will describe the use of the proposed stopping rule subsequently.

To construct Markov kernels invariant with respect to each distribution $\tilde{\pi}_{\lambda}$, we used Gibbs kernels constructed in the manner of Algorithm~\ref{alg:gcmcMCMC}.
That is to say that in each time step of the SMC sampler (i.e.\ for each value of $\lambda$) and for each particle, each of $X_{1:b}$ was updated by drawing exactly from its conditional distribution, after which $Z$ was updated similarly.

Figure~\ref{fig:smcNormal_all} shows the SMC estimate $\pi_{\lambda}^{N}(\Id)$ obtained for each $\lambda$, in a single run of this algorithm.
To determine a subset of these estimates to be used for local linear regression, we used the approach described in Section~\ref{subsec:inclusionProcedure}; this subset is displayed in Figure~\ref{fig:smcNormal_subset}.
In this case, we see that for the smallest values of $\lambda$ considered, the estimates exhibit increased variance, due to the poorer mixing of the Markov kernels, and the degeneracy of the particle set.

\begin{figure}
	\centering
	\includegraphics[width=.6\linewidth]{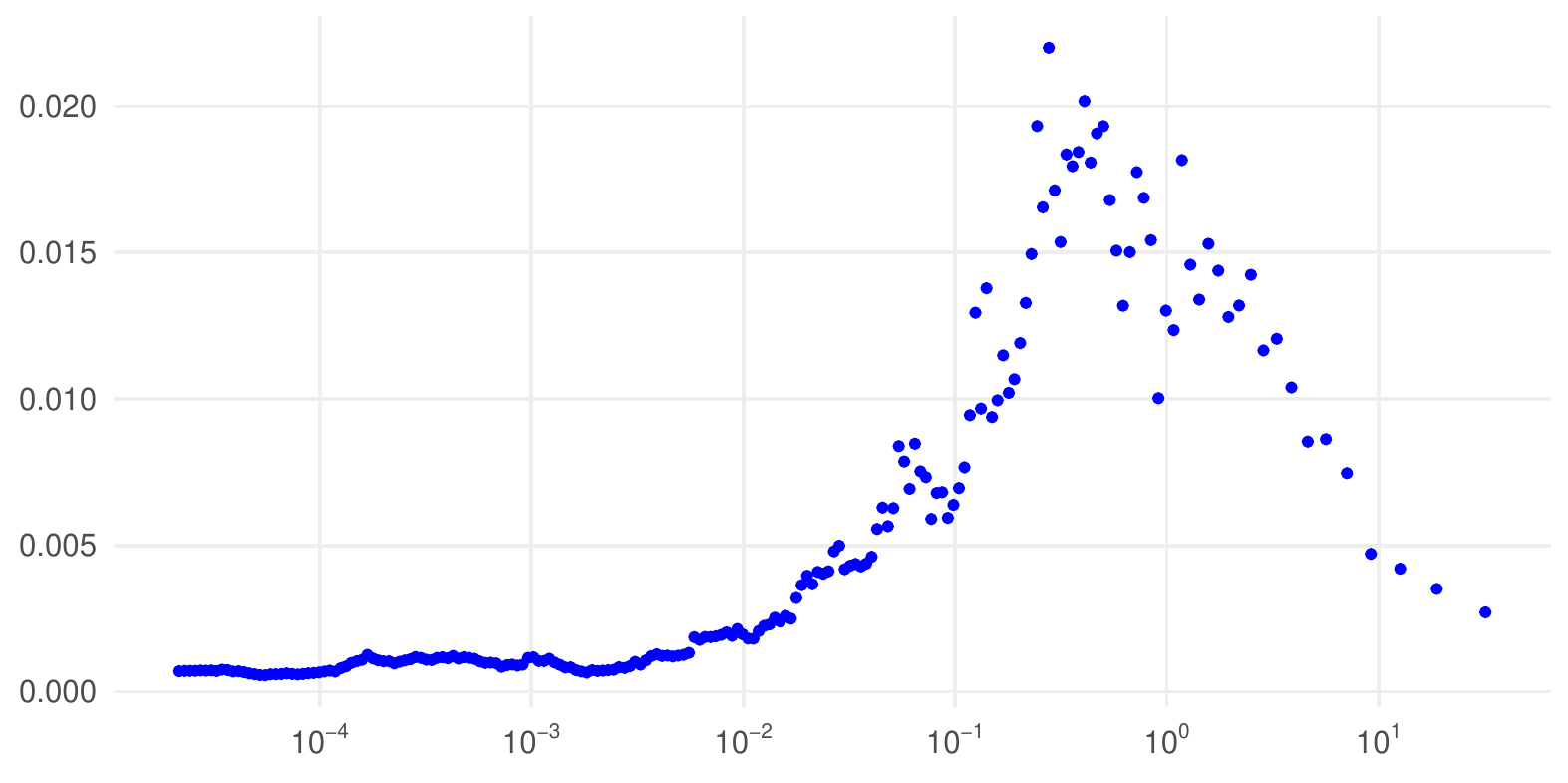}
	\caption{For the estimates $\pi_{\lambda}^{N}(\Id)$ plotted in Figure~\ref{fig:smcNormal_subset}, the estimates' relative weights as used in the weighted least squares bias correction technique, plotted against $\lambda$ on a logarithmic scale. }
	\label{fig:smcNormalWeights}
\end{figure}

As described in Section~\ref{subsec:varianceEstimation}, when conducting local least squared regression we weight each estimate in inverse proportion to its estimated (asymptotic) variance.
For the estimates plotted in Figure~\ref{fig:smcNormal_subset}, these relative weights are presented in Figure~\ref{fig:smcNormalWeights}, with $\lambda$ on a log scale for clarity.
The resulting weighted least squares fit is overplotted in Figure~\ref{fig:smcNormal_subset}, together with the corresponding \emph{unweighted} (ordinary) least squares fit.
We see that for these results, the weighted least squares fit better reflects the local linear dependence on $\lambda$, being less influenced by the high-variance estimates near $0$, which correspondingly carry less weight in the regression.

As discussed in Section~\ref{sec:SMCimplementation}, we may view the SMC sampler as a method to improve or `refine' the estimator that would be formed using the initial set of particles, i.e.\ $\pi_{\lambda_{0}}^{N}(\Id)$, where $\lambda_{0} = 1000$.
A straightforward choice of such a refined estimator would therefore be the SMC estimate $\pi_{\lambda_{n}}^{N}(\Id)$ corresponding to $\lambda_{n}$, the final (smallest) $\lambda$ value considered.
We ran the SMC sampler $25$ times; the value of $\lambda_{n}$ varied between runs due to the adaptive specification of the sequence of distributions, but each time was approximately $2.2 \times 10^{-5}$.
The mean squared error of the resulting estimate, computed over the $25$ replicates, was $1.13 \times 10^{-3}$.
This is rather lower than the MSE of the estimator $\pi_{\lambda_{0}}^{N}(\Id)$ corresponding to the initial value $\lambda_{0} = 1000$, which was $1.32 \times 10^{-2}$.

For each run of the SMC sampler we also computed a bias-corrected estimate \eqref{eq:biasCorrectedEstimate} of $\pi(\Id)$ using weighted least squares as described above; that is, the intercept of the local least squares linear fit.
The mean squared error of this bias-corrected estimator was $3.60 \times 10^{-5}$, rather lower MSE than the simpler approach of considering solely the final $\lambda$ value.
Additionally, for purposes of comparison we also computed a bias-corrected estimate using ordinary (unweighted) least squares.
This resulted in a mean squared error of $2.57 \times 10^{-4}$, between the values from the other approaches.

We subsequently considered the effects of using the stopping rule proposed in Section~\ref{subsec:stoppingRule}, retrospectively applying the procedure of Algorithm~\ref{alg:gcmcSMCstoppingRule} with $\kappa = 15$ to each of the $25$ simulations.
On average, termination occurred after $53.0$ SMC iterations (i.e.\ values of $\lambda$ following the initial $\lambda_{0}$).
The SMC estimate chosen by the stopping rule, estimated to have the lowest mean squared error, was found to have an MSE of $1.11 \times 10^{-5}$, lower than the values for which $n=200$ iterations were used.
The final bias-corrected estimate at the time of termination was found to have a comparably low MSE of $9.23 \times 10^{-6}$.

We see therefore that using the stopping rule to choose the SMC estimate of lowest estimated MSE here results in superior estimator to that obtained by simply taking the final SMC estimate after a fixed number of iterations, representing a significant refinement of the estimator formed using the initial particle set.
We also find that the bias-corrected estimate here performs slightly better than the corresponding estimate obtained after using $200$ iterations: in that case, the SMC estimates corresponding to the very smallest values of $\lambda$ are of high variance and can distort the regression, despite being appropriately weighted.

\subsection{Log-normal example}
\label{subsec:additionalLognormalExample}

We present an additional example based on the log-normal model presented in Section~\ref{subsec:lognormalExample} of the main manuscript.
Again, we compare Global Consensus Monte Carlo (GCMC), using the Gibbs sampler form of Algorithm~\ref{alg:gcmcMCMC}, with three embarrassingly parallel methods based on subposterior sampling: the Consensus Monte Carlo (CMC) method of \citet{ScottEtAl2016}, the nonparametric density product estimation (NDPE) approach of \citet{NeiswangerEtAl2014}, and the Weierstrass rejection sampling (WRS) combination technique of \citet{WangDunson2013}.

We generated a data set comprising $b = 32$ blocks, each containing $10^{4}$ data.
Within the $j$th block, the data were generated as i.i.d.\ observations of a log-normal random variable with parameters $(\mu_j, 1)$; the parameters $\mu_j$ were drawn independently from a normal distribution with mean $0$ and variance $10^{-2}$.
We took $f_{j}(z)=\dlnorm(\bar{y}_{j};\log(z),10^{-4})$, with each $\bar{y}_{j}$ being the geometric mean of the observations in the $j$th block; we used the same prior $\mu(z)=\dlnorm(z;0,25)$ as previously.
While this represents a misspecified model, it is useful in exemplifying the behaviour of global consensus Monte Carlo in cases where there are differences between the blocks of data.

\begin{table}
	\begin{centering}
		\begin{tabular}{llcc}
			\multicolumn{2}{l}{}
			& $\varphi(z) = z$
			& $\varphi(z) = \log(z)$
            \\
			\multicolumn{2}{l}{Algorithm}
			& ($\pi(\varphi) = 1.01139$)
			& ($\pi(\varphi) = 0.01132$)
            \\
			\hline
			\multirow[t]{6}{*}{GCMC} & $\lambda=10^{1}$
			& $1.18022 \pm 0.002097$
			& $\hphantom{-}0.01121 \pm 0.001537$
            \\
			& $\lambda=10^{0}$
			& $1.02740 \pm 0.000609$
			& $\hphantom{-}0.01142 \pm 0.000623$
            \\
			& $\lambda=10^{-1}$
			& $1.01305 \pm 0.000156$
			& $\hphantom{-}0.01140 \pm 0.000154$
            \\
			& $\lambda=10^{-2}$
			& $1.01155 \pm 0.000067$
			& $\hphantom{-}0.01133 \pm 0.000066$
            \\
			& $\lambda=10^{-3}$
			& $1.01140 \pm 0.000020$
			& $\hphantom{-}0.01132 \pm 0.000020$
            \\
			& $\lambda=10^{-4}$
			& $\mathbf{1.01139 \pm 0.000013}$
			& $\mathbf{\hphantom{-}0.01133 \pm 0.000013}$
            \\
			& $\lambda=10^{-5}$
			& $1.01139 \pm 0.000023$
			& $\hphantom{-}0.01132 \pm 0.000023$
            \\
            \hline
			\multirow[t]{2}{*}{CMC}
			& data not permuted
			& $0.99828 \pm 0.000081$
			& $-0.00172 \pm 0.000081$
            \\
			& data permuted
			& $1.01141 \pm 0.000007$
			& $\hphantom{-}0.01135 \pm 0.000007$
            \\
            \hline
            \multirow[t]{2}{*}{NDPE}
            & data not permuted
            & $1.01556 \pm 0.000077$
            & $\hphantom{-}0.01518 \pm 0.000076$
            \\
            & data permuted
            & $1.01155 \pm 0.000077$
            & $\hphantom{-}0.01123 \pm 0.000077$
            \\
            \hline
            \multirow[t]{2}{*}{WRS}
            & data not permuted
            & $0.99867 \pm 0.000420$
            & $-0.00133 \pm 0.000420$
            \\
            & data permuted
            & $1.01135 \pm 0.000039$
            & $\hphantom{-}0.01129 \pm 0.000039$
            \\
			\hline
		\end{tabular}
		\par\end{centering}
	\caption{
		True values and estimates of $\pi(\varphi)$, for various test functions $\varphi$, for the log-normal toy model described in this supplement.
        For the three embarrassingly parallel approaches (CMC, NDPE, WRS) we present results obtained both without and with first permuting and repartitioning the data into new blocks.
		For each method the mean estimate $\pm$ Monte Carlo standard error is presented, as computed over $25$ replicates; the estimator corresponding to the lowest mean squared error is printed in bold.
	}
	\label{tab:lognormalResults2}
\end{table}

Table~\ref{tab:lognormalResults2} shows the estimates of $\int z \pi(z) \diff z$ and $\int \log(z) \pi(z) \diff z$, from $25$ runs in each algorithmic setting.
Global consensus Monte Carlo produces low-bias estimates for a range of $\lambda$ values.
In contrast, the embarrassingly parallel methods result in somewhat larger biases; this is particularly the case for the expected value of the logarithm in the cases of CMC and WRS, which behave similarly on this example.
The NDPE method, which is based on kernel density estimation, works reasonably well for this univariate model.

When the data are first randomly permuted and repartitioned into $32$ new blocks, the performances of the embarrassingly parallel methods are improved, though we still find that for appropriately-chosen $\lambda$, GCMC estimators attain a lower mean squared error.
Furthermore, for large distributed data sets permutation of the data in this manner may not be feasible, for example if security restrictions prevent the transfer of data between machines.

\bibliographystyle{apalike}

\bibliography{refs}